\documentclass{aa}  

\usepackage{graphicx}
\usepackage{txfonts}
\usepackage{hyperref}
\hypersetup{
    colorlinks=true,
    linkcolor=cyan,
    citecolor=cyan,
    urlcolor=cyan,
    }
\usepackage{booktabs}
\begin{document}

   \title{Search for new Galactic Wolf-Rayet stars using \textit{Gaia} DR3}

   \subtitle{I. Candidate selection and the follow-up of the bright sample}

   \author{Lionel Mulato\inst{1}
   \and
   Jaroslav Merc
          \inst{2,3}
          \and
          Stéphane Charbonnel\inst{1}
          \and
          Olivier Garde\inst{1}
          \and
          Pascal Le Dû\inst{1}
          \and
          Thomas Petit\inst{1}
          }

   \institute{Southern Spectroscopic Project Observatory Team (2SPOT), 45, Chemin du Lac 38690 Châbons, France\\
   \email{lionelmulato@gmail.com}
         \and
             Astronomical Institute of Charles University, V Hole\v{s}ovi\v{c}k{\'a}ch 2, Prague, 18000, Czech Republic
             \\
   \email{jaroslav.merc@mff.cuni.cz}
            \and
            Instituto de Astrof\'isica de Canarias, Calle Vía Láctea, s/n, E-38205 La Laguna, Tenerife, Spain
             }

   \date{Received December 09, 2024; accepted February 23, 2025}

% \abstract{}{}{}{}{} 
% 5 {} token are mandatory
 
  \abstract
   {\textit{Gaia} DR3, released in June 2022, included low-resolution XP spectra that have been used for the classification of various types of emission-line objects through machine-learning techniques. The \textit{Gaia} \emph{Extended Stellar Parametrizer for Emission-Line Stars} (ESP-ELS) algorithm identified 565 sources as potential Wolf-Rayet (WR) stars. Over half of them were already known as WR stars in the Milky Way and Magellanic Clouds.}
   {This study aimed to utilize \textit{Gaia} DR3 data to identify new Galactic WR stars. }
   {We extracted all sources classified as WC or WN type stars by the ESP-ELS algorithm from the \textit{Gaia} catalog. By applying judicious 2MASS color selection criteria, leveraging \textit{Gaia} H$\alpha$ measurements, and filtering out objects already cataloged in various databases, we selected 37 bright candidates ($G \leq $ 16 mag) and 22 faint candidates ($G > $ 16 mag). Spectroscopic follow-up observations of these candidates were conducted using the 2SPOT facilities in Chile and France, as well as 1-m C2PU's Epsilon telescope at the Calern Observatory.}
   {This paper focuses on the brighter sample. Among the 37 targets, we confirmed 17 and 16 new Galactic WC and WN type WR stars, respectively. Three of them were recently reported as new WR stars in an independent study.}
   {The \textit{Gaia} mission provides a valuable resource for uncovering WR stars missed in earlier surveys. While this work concentrated on a relatively small starting sample provided by the ESP-ELS algorithm, our findings highlight the potential for refining selection criteria to identify additional candidates not included in the outputs of the algorithm. Furthermore, the observation program underscores the utility of small telescopes in acquiring initial spectral data for sources with magnitudes up to $G \sim 16$ mag.}

   \keywords{Stars: Wolf-Rayet -- Stars: massive -- Stars: early-type
               }

   \maketitle

%-------------------------------------------------------------------

\section{Introduction}

   \citet{1867CRAS...65..292W} reported the discovery of three stars exhibiting strong emission lines of unknown origin: HD 191765, HD 192103, and HD 192641. These stars became prototypes for the Wolf-Rayet (WR) class, and their optical appearance served as a reference for subsequent discoveries \citep{1996LIACo..33....1V}. It is now established that Population I WR stars represent a brief, terminal phase in the evolution of the most massive O-type stars (with initial masses > 25 $M_\odot$). They are progenitors of core-collapse supernovae \citep{2007ARA&A..45..177C} and black holes (BH), playing an important role in BH-BH mergers \citep[e.g.,][]{2010ApJ...725..816B}, given that a large fraction of WR stars \citep[>40 -- 50\%;][]{2001NewAR..45..135V,2024arXiv241004436S} live in binary systems. Furthermore, WR stars significantly influence the interstellar medium (ISM) through high mass-loss rates and intense radiation, impacting star formation and contributing to ISM chemical enrichment.
   
   At their current evolutionary stage, the Population I WR stars are highly evolved hot (>30 kK), still massive (10-25 $M_\odot$), and luminous (>$10^5$ $L_\odot$) stars, having lost their outer hydrogen-rich envelopes \citep{2007ARA&A..45..177C, 2024arXiv241004436S}. Their dense, fast stellar winds of ionized heavy elements define the WR phenomenon \citep{2001NewAR..45..135V}, characterized by spectra dominated by broad emission lines of highly ionized helium, nitrogen, carbon, and oxygen, along with a hot O-type continuum. WR stars are classified into three types: WN dominated by helium and nitrogen lines \citep{1996MNRAS.281..163S}, WC with prominent helium, carbon, and oxygen lines, and the rare WO whose spectra are similar to that of the WC-type but with stronger oxygen lines \citep{1998MNRAS.296..367C}.

   The population of WR stars in the Milky Way is estimated to be around 1200 \citep{2015MNRAS.449.2436R} -- 6000 \citep{2009AJ....138..402S}, yet only 676 are cataloged in the online Galactic WR catalog (GWRC; v1.29; May 2024)\footnote{\hyperlink{https://pacrowther.staff.shef.ac.uk/WRcat/}
   {https://pacrowther.staff.shef.ac.uk/WRcat/}} \citep{2015wrs..conf...21C, 2015MNRAS.447.2322R}. It can be explained by the relatively short lifetimes of WR progenitors (a few Myr), which confine them to their birth sites within dusty molecular clouds \citep{2011AJ....142...40M}. WR stars themselves contribute to dust production \citep{1987QJRAS..28..248W, 1991MNRAS.252...49U, 2022NatAs...6.1308L}, and dust extinction complicates optical detection at large distances despite their high luminosity. Infrared surveys have significantly advanced WR star detection over the past two decades, employing techniques such as near-IR imaging with interference filters and color-based selection from 2MASS or GLIMPSE catalogs. These methods have uncovered several hundred new WR stars, with spectroscopic follow-up typically conducted in the near-IR range \citep[e.g.,][]{2009AJ....138..402S, 2012AJ....143..149S, 2015MNRAS.452.2858K, 2011AJ....142...40M, 2012AJ....144..166S, 2018MNRAS.473.2853R}.
   
   The WR phenomenon is also observed among some central stars of planetary nebulae (CSPNe). These low-mass Population II WR stars are denoted as [WC], [WO], or [WN], with brackets distinguishing them from massive Population I WR stars. Of the $\sim$130 [WR] stars cataloged in \cite{2020A&A...640A..10W}, most are [WC] stars, while [WN] and [WO] types are rare \citep{2000A&A...364..597G, 2001A&A...370..513G, 2003MNRAS.346..719M, 2003A&A...403..659A, 2010PASA...27..129F}. WR and [WR] stars of the same subclass exhibit similar spectra, and additional information, such as distance and absolute magnitude, is necessary to discriminate between them  \citep{2001A&A...370..513G}.
    
    This work focuses on massive WR stars. We first briefly analyze the sample of known Population I WR stars that have counterparts in the \textit{Gaia} DR3 main catalog \citep{2016A&A...595A...1G,2023A&A...674A...1G}. In Section \ref{sec:Massive Wolf Rayet stars in GAIA DR3 data} we provide a statistical overview of their number and classes. 
    Section \ref{sec:Our searching method} describes our methodology for identifying strong WR candidates in \textit{Gaia} DR3. We detail the selection criteria applied to the output of the ESP-ELS algorithm. We conducted a spectroscopic follow-up of candidates, focusing on the brighter end of the distribution ($G \leq $ 16 mag). The observations carried out mainly with the small amateur telescopes are presented in Section \ref{sec:Spectroscopic follow-up}. 
    Results for all 37 bright candidates with $G \leq 16$ mag are presented in Section \ref{sec:Results and spectral classification}, along with a discussion of their classification as Population I massive WR stars or Population II low-mass [WR] stars in Section \ref{sec:Population I massive WR stars or Population II low mass [WR]}. Finally, Section \ref{sec:Conclusion} summarizes our findings and highlights some inconsistencies in positions and identifications of WR stars in online databases.

\begin{figure}[t]
   \centering
   \includegraphics [width=\columnwidth] {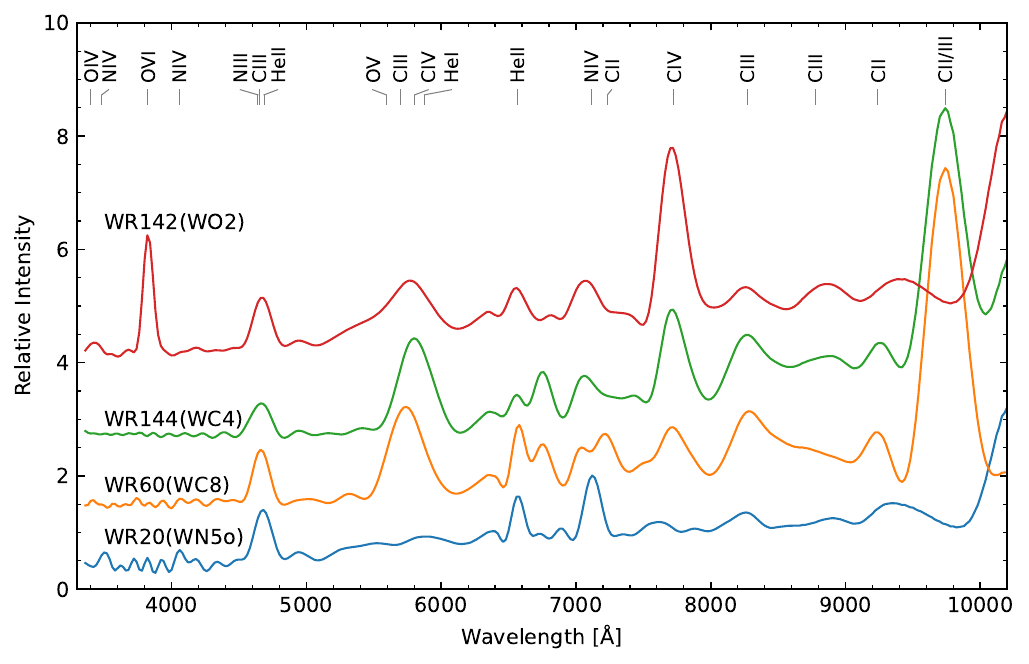}
   \caption{Typical \textit{Gaia} XP spectra of WR stars with main helium, nitrogen, carbon and oxygen lines marked.}
   \label{fig:GAIA_XPSpectra_of_WRs}%
\end{figure}

%--------------------------------------------------------------------
\section{Massive Wolf Rayet stars in \textit{Gaia} DR3 data}
\label{sec:Massive Wolf Rayet stars in GAIA DR3 data}

We cross-matched 676 known Population I WR stars from the GWRC (v1.29; May 2024) with the \textit{Gaia} DR3 main catalog using a 1.5-arcsecond search radius. 
This process identified 443 known WR stars with counterparts in \textit{Gaia} DR3. Most WR stars absent from the main catalog are faint sources ($J > 14$ mag) that were primarily discovered through the infrared detection methods previously discussed.

Approximately 60\% of all known WR stars are of the WN type, while 40\% are WC type, with the \textit{Gaia}-detected sample reflecting a similar distribution. Of the four known WO stars, WR30a, WR93b, WR102, and WR142, all have counterparts in \textit{Gaia}. Figure \ref{fig:GAIA_XPSpectra_of_WRs} illustrates typical \textit{Gaia} XP spectra \citep{2023A&A...674A...2D,2023A&A...674A...3M} for WN, WC, and WO stars, with the main spectral features of each subclass clearly visible despite the low resolution of the data.

The XP spectra were utilized by the \emph{Extended Stellar Parametrizer for Emission-Line Stars} (ESP-ELS) algorithm \citep{2023A&A...674A..26C,2023A&A...674A..28F} to identify and classify 57\,511 emission-line stars (ELS) of various types, including WR stars. The classification model was trained on seven ELS categories: WC, WN, Be stars, Herbig Ae/Be stars, T Tauri stars, active M dwarfs, and PNe.

The ESP-ELS algorithm classified 229 of the 443 known Galactic WR stars in the \textit{Gaia} DR3 main catalog as WR stars, successfully assigning subclasses (WC or WN) to all but one star, WR 120-1. Unfortunately, no sampled nor continuous XP spectrum is available for WR 120-1 (WC9), precluding further investigation into the reasons for this misclassification. The WN and WC subclass distribution in the ESP-ELS classifications closely matches that of the original Galactic WR star sample, demonstrating the comparable effectiveness of the algorithm in identifying both subclasses.

\begin{figure}[t]
   \centering
   \includegraphics[width=0.49\textwidth]{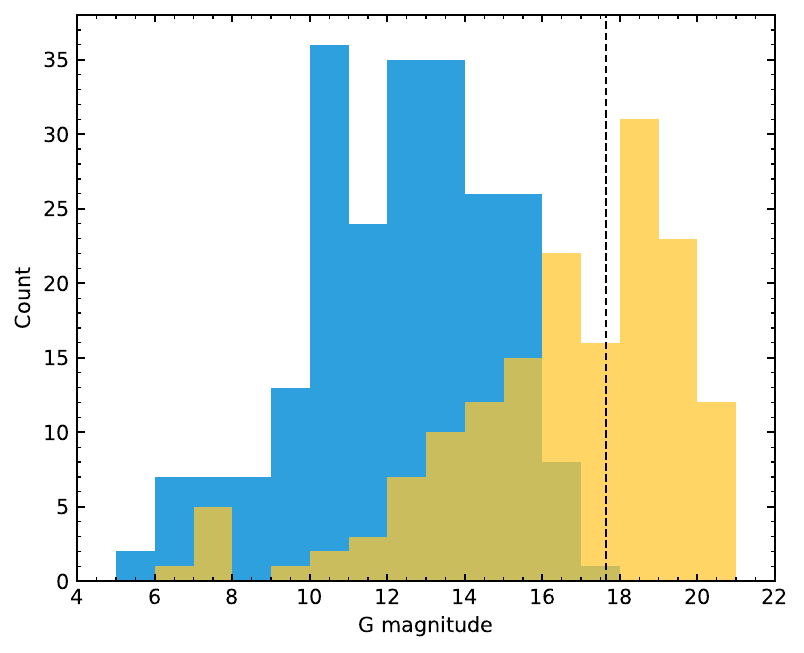}
   \caption{Distribution of the \textit{G} magnitudes of known Galactic WR stars. Stars included in the \textit{Gaia} ESP-ELS analysis are shown in blue, while the objects not analyzed by the algorithm (for various reasons; see the text for details) are shown in yellow. The dashed black line represents the processing limit of the algorithm ($G <$ 17.65 mag).}
   \label{fig:Known_WR_detected_by_ELS}%
\end{figure}

Conversely, the ESP-ELS algorithm failed to classify 214 of the 443 WR stars with counterparts in the \textit{Gaia} DR3 main catalog as WR stars. As shown in Fig. \ref{fig:Known_WR_detected_by_ELS}, 73 of these stars have $G$ magnitudes exceeding the processing limit of the ESP-ELS algorithm ($G$ = 17.65 mag), and an additional 54 lack $G$ measurements. However, the algorithm also missed several dozen bright sources ($G \leq 16$ mag). Most pre-calibrated XP spectra for these sources, available in the \textit{Gaia} main catalog, appear featureless, exhibiting unusual continua and lacking prominent carbon or nitrogen lines. As a result, the algorithm could not assign classifications to most of these stars. Only seven of these sources were assigned alternative ELS classifications, such as Herbig-Haro objects, T Tauri stars, or PNe, with non-zero probabilities (10–45\%) of being WC or WN stars. 

Interestingly, three of the four known WO stars were classified as PNe by the ESP-ELS algorithm, with WR30a being the only WO star unassigned an ELS classification. This result is surprising, as WO stars do not typically exhibit prominent PN spectral features, such as strong emission in the 5000 Å region caused by the blending of the [\ion{O}{iii}] $\lambda$$\lambda$4959, 5007 doublet or H$\beta$ emission. One might have expected these WO stars to be classified as WC stars since both subclasses share strong emission features in the \ion{C}{iv} $\lambda$5804 and \ion{C}{iii} $\lambda$4650 regions. However, the wavelength ranges used to train the ESP-ELS algorithm do not include the \ion{C}{iv} $\lambda$5804 line \citep[see the \textit{Gaia} DR3 online documentation;][]{2022gdr3.reptE..11U,2022gdr3.reptE....V}. This likely contributed to the misclassification, as the near-UV and near-IR lines characteristic of WO stars may have been confused with similar emissions from PNe.

%--------------------------------------------------------------------
\section{Search for new WR stars}
\label{sec:Our searching method}
Our search for new WR stars began with the serendipitous discovery of a WR star during a search for new symbiotic systems using \textit{Gaia} DR3 data. The method developed for identifying symbiotic stars has proven highly effective, leading to the discovery of dozens of new symbiotic systems (Merc et al., in prep.). As a byproduct, one of our symbiotic candidates turned out to be a new WR star, denoted WR-C-14 in this work (see Table \ref{tab:WR_Candidates_list}). %which is the first massive WR star discovered by an amateur setup.
Encouraged by this discovery, we refined our selection criteria (outlined below) to specifically target new WR stars.

\subsection{Selection criteria}

To identify the most promising candidates for spectroscopic follow-up, we extracted all sources classified as WC or WN stars by the ESP-ELS algorithm from \textit{Gaia} DR3. This resulted in a sample of 565 WR candidates: 136 classified as WC and 429 as WN. Additionally, two more WN candidates were mentioned in \citet{2023A&A...674A..28F} but were apparently not included in the final \textit{Gaia} DR3 table.

\begin{figure}[!h]
   \centering
   \includegraphics[width=\columnwidth]{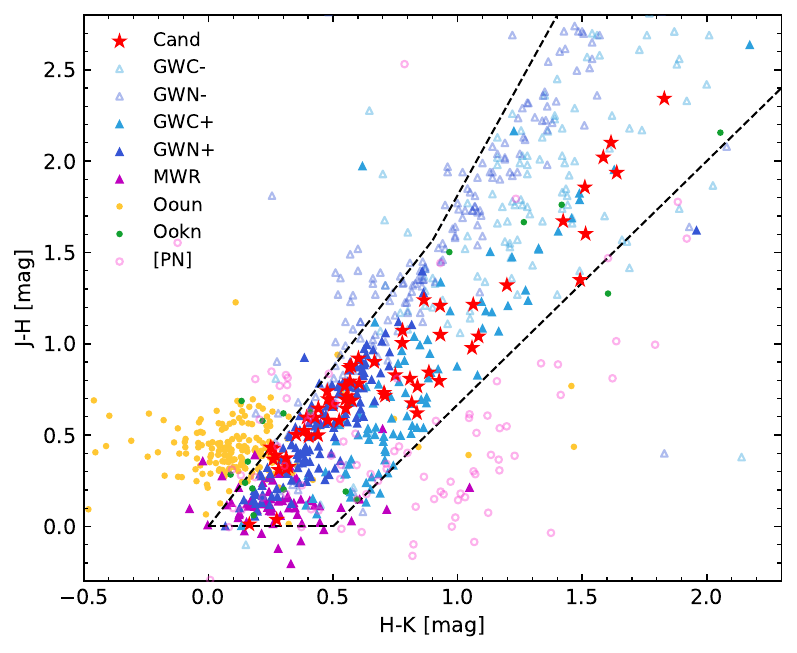}
   \hspace{0.005\textwidth}
   \includegraphics[width=\columnwidth]{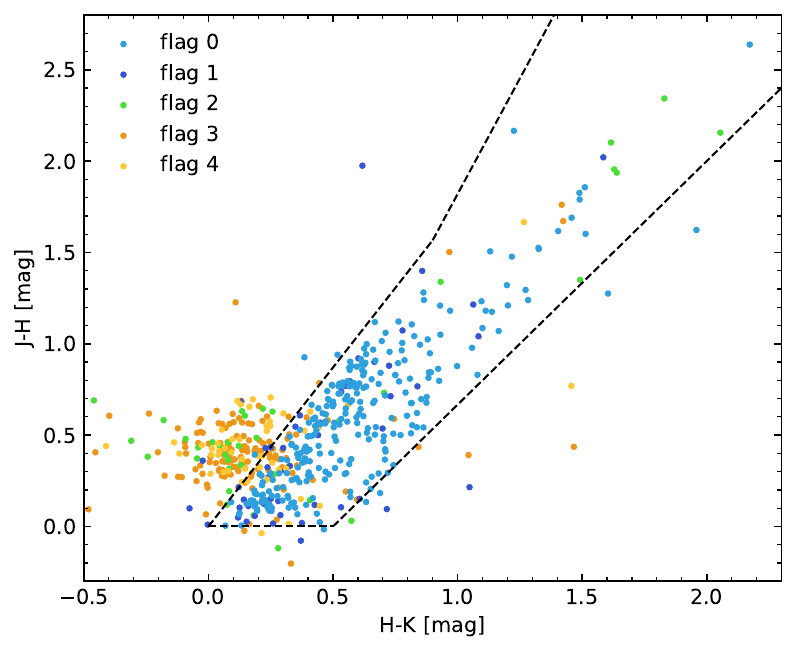}
   \caption{Position of WR stars and candidates in the 2MASS color-color diagram. Solid black lines define the near-IR region of WR stars \citep{2011AJ....142...40M}. \textbf{Upper panel}: Sources identified as WR stars by \textit{Gaia} ESP-ELS algorithm are shown as color-filled markers, and are divided into several groups based on the analysis presented in this paper (see the text): previously known Galactic WC (GWC+; light blue triangles) and WN stars (GWN+; dark blue triangles), known WR stars in the Magellanic Clouds (MWR; purple triangles), our selected candidates (Cand; red star symbols), and the remaining objects with known (Ookn; green circles) and unknown nature in the SIMBAD database (Ooun; yellow circles). Known WR and [WR] stars not classified by the algorithm are shown as empty markers: known Galactic WC (GWC-; light blue triangles) and WN (GWN-; dark blue triangles) stars, and [WR] CSPNe ([PN]; pink circles). \textbf{Bottom panel}: Position of the sources identified as WR stars by the \textit{Gaia} ESP-ELS algorithm, color-coded by the quality flag. The lower the value of the flag, the higher the probability that the source is a WR star.}
   \label{fig:WR_NIR_colors_diagram}%
\end{figure}

As an initial step, we excluded 229 known Galactic WR stars previously discussed. Next, we filtered out 188 sources with near-IR colors atypical for WR stars. Figure~\ref{fig:WR_NIR_colors_diagram} illustrates the position of all ESP-ELS WR sources in a near-IR color-color diagram constructed using 2MASS observations \citep{2006AJ....131.1163S}, alongside the typical WR near-IR region \citep{2011AJ....142...40M}. These non-WR sources cluster in specific regions of the diagram, are discussed further in Sect. \ref{subsec:Rejected sources with non WR NIR colors}. The diagram also includes known WR stars not identified by the ESP-ELS algorithm and [WR] CSPNe from \cite{2020A&A...640A..10W}. Known WR stars undetected by the algorithm exhibit higher average \(J-H\) and \(H-K\) values, likely due to greater dust extinction. Conversely, the [WR] CSPNe occupy distinct regions, mostly outside the typical WR near-IR domain.

\begin{figure}[!h]
   \centering
   \includegraphics[width=0.49\textwidth]{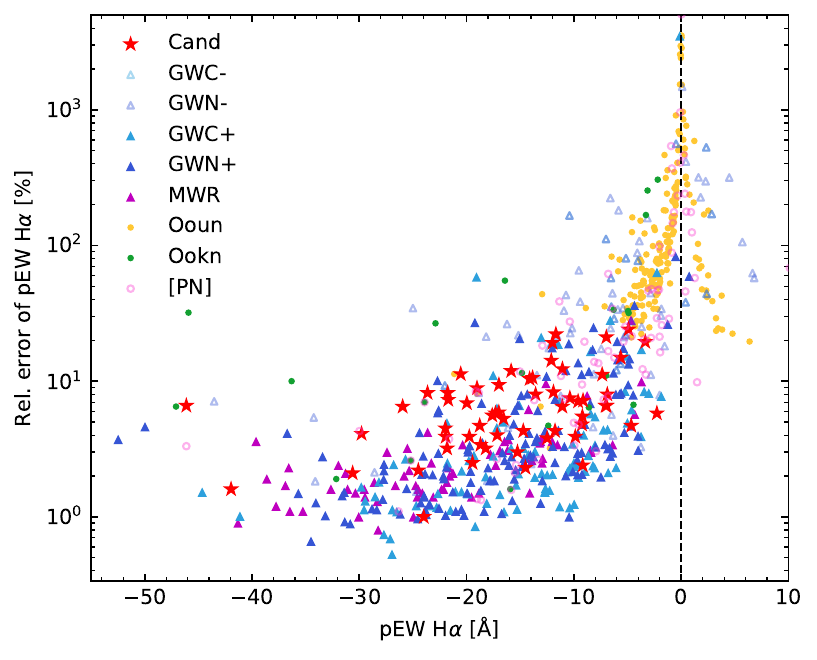}
   \hspace{0.005\textwidth}
   \includegraphics[width=\columnwidth]{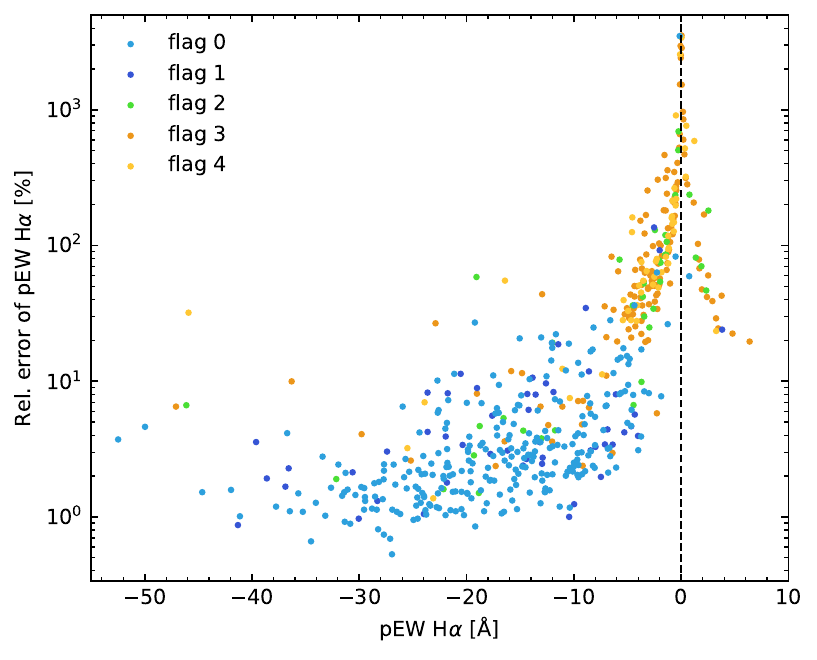}
   \caption{\textit{Gaia} pseudo-equivalent width of the H$\alpha$ line. \textbf{Upper panel}: Position of sources identified as WR stars by the \textit{Gaia} ESP-ELS algorithm (color-filled markers), along with known Galactic WR stars not classified by the algorithm and [WR] CSPNe (edge-colored markers). The symbols are the same as in the Fig. \ref{fig:WR_NIR_colors_diagram}. \textbf{Bottom panel}: Position of the sources identified as WR stars by the \textit{Gaia} ESP-ELS algorithm, color-coded by the quality flag.}
   \label{fig:Ewha_rel_err}%
\end{figure}

To ensure sample quality, we also incorporated \textit{Gaia} H$\alpha$ pseudo-equivalent width measurements (pEW H$\alpha$), which are effective for identifying true H$\alpha$ emitters. For WR stars, the emission detected by \textit{Gaia} in most cases arises not from hydrogen but from broad helium or carbon emission lines coinciding with the H$\alpha$ region. Leveraging this overlap, we retained only sources with negative pEW H$\alpha$ values (indicative of emission) and a relative error below 30\%. This filtering removed 8 sources with positive pEW H$\alpha$ and 20 with high relative errors. These criteria are motivated by the distribution of pEW H$\alpha$ and its relative error for known WR stars, detected and undetected as WR stars by the ESP-ELS algorithm, as well as [WR] CSPNe, as shown in Fig.~\ref{fig:Ewha_rel_err}. 

\subsection{Cross-match with SIMBAD}

Finally, we cross-checked the remaining candidates with the SIMBAD database \citep{2000A&AS..143....9W}, removing all objects of confirmed nature. This included 51 WR stars in the Magellanic Clouds (not included in the Galactic WR database used for crossmatching), two cataclysmic variables, one Herbig-Haro object, one high-mass X-ray binary, a star named HD 45166 \citep[identified as a WR star in SIMBAD but absent from the GWRC -- given the peculiar nature it is often classified as a quasi-WR and is probably a post-merger of two lower-mass helium stars; see][]{2023Sci...381..761S}, and five additional sources discussed in more details below: one WR candidate (IPHAS J195935.55+283830.3), one massive WR star (WR31-1), and three [WR] PNe.

IPHAS J195935.55+283830.3 is a WR candidate with an optical spectrum available in \cite{2010A&A...509A..41C}. Its near-IR colors are consistent with those of WR stars, making it a compelling candidate. The authors also reported a faint, asymmetric, and very extended nebula near the star, with the target positioned close to the bright, bow-shaped eastern edge. This nebula was proposed as a possible WR ring nebula. In 2019, our group obtained a low-resolution spectrum of this bright arc, now available in the HASH PN database\footnote{\hyperlink{http://202.189.117.101:8999/gpne/index.php}{http://202.189.117.101:8999/gpne/index.php}} \citep{2017IAUS..323..327B}. The spectrum shows shock-excitation features, including strong [\ion{S}{ii}] $\lambda$$\lambda$6716-31, moderately strong [\ion{N}{ii}] $\lambda$$\lambda$6548-83, and [\ion{O}{i}] $\lambda$6300, along with high excitation 
 [\ion{O}{iii}] $\lambda$$\lambda$4959, 5007 emission stronger than H$\beta$. These features are consistent with a WR ring nebula formed by stellar winds but could also be attributed to an ancient supernova remnant.

\begin{figure}[]
   \centering
   \includegraphics[width=\columnwidth]{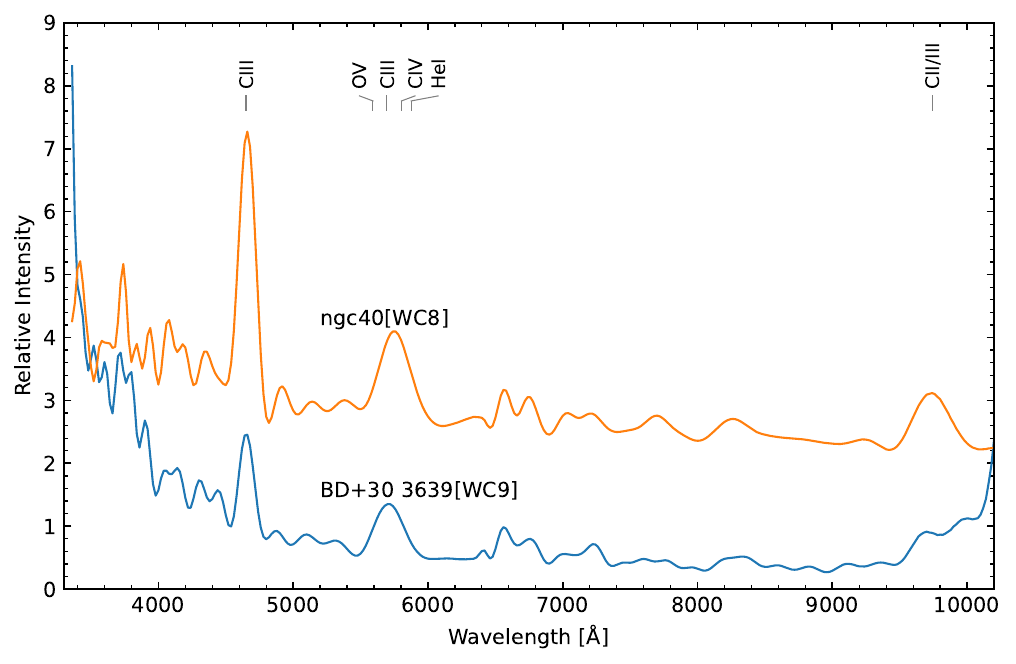}
   \caption{\textit{Gaia} XP spectra of the [WR] PNe NGC 40 and BD+30 3639. Both spectra show typical [WC] features in their spectrum. NGC 40 is classified as a PN by the ESP-ELS Algorithm, while BD+30 3639 has no assigned ELS class.}
   \label{fig:GAIA_XPSpectra_of_[WR]}%
\end{figure}

WR31-1 (THA 35-II-153) was initially missed during our crossmatch with the online GWRC because the database provides incorrect coordinates for this object. The correct position is listed in SIMBAD.

The three [WR] CSPNe excluded from our final list are PMR 1, NGC 6751, and RaMul 2. PMR 1 and NGC 6751 are listed in the CSPNe catalog \citep{2020A&A...640A..10W}, while RaMul 2 was recently discovered, spectroscopically confirmed by amateur astronomers \citep{2022A&A...666A.152L}, and further analyzed by \cite{2024A&A...686A..29W}.

The CSPNe catalog by \cite{2020A&A...640A..10W} includes approximately 130 [WR] stars. Of these, only PMR 1 and NGC 6751 were classified as WN stars by the ESP-ELS algorithm (RaMul 2 is not part of the catalog). Notably, more than 75\% of the [WR] CSPNe in the catalog lack an assigned ELS class, while the remaining 25\% are classified as PNe due to contamination from nebular lines. Pre-calibrated XP spectra for 27 [WR] CSPNe are available in the \textit{Gaia} main catalog.

Figure \ref{fig:GAIA_XPSpectra_of_[WR]} presents representative XP spectra for two [WR] CSPNe: NGC 40 ([WC8]) and BD+30 3639 ([WC9]). The characteristic features of the [WC] subclass are clearly identifiable in both spectra. However, NGC 40 is classified as a PN by the ESP-ELS algorithm, while BD+30 3639 remains unclassified.

\subsection{Sources with non-WR near-IR colors}
\label{subsec:Rejected sources with non WR NIR colors}

We examined the 188 sources rejected due to their non-WR near-IR colors in more detail. A crossmatch with the SIMBAD database revealed 16 known Magellanic WR stars and one additional Galactic WR star, WR111-6. This source was missed during the initial crossmatch with the online GWRC because the database provides incorrect coordinates \citep[taken from the discovery paper of ][interestingly, they did not provide a finding chart for this particular object]{2012AJ....143..149S}. It should correspond to the coordinates of Gaia DR3 4095702533139975168.

\begin{figure}[]
   \centering
   \includegraphics[width=\columnwidth]{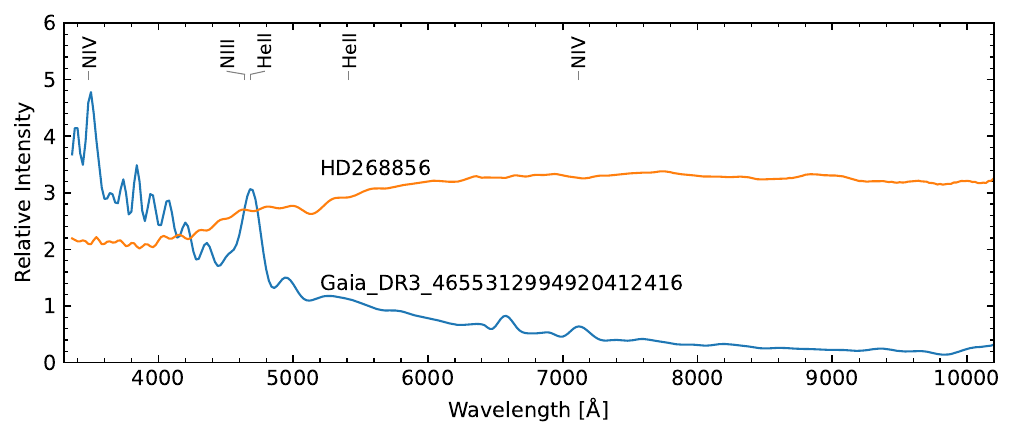}
   \caption{\textit{Gaia} XP spectra of Gaia DR3 4655312994920412416 and HD 268856. The correct counterpart of the Magellanic WR star Brey 11  is Gaia DR3 4655312994920412416.}
   \label{fig:HD268856_false_WR}%
\end{figure}

Among the remaining sources, only 9 are known objects in SIMBAD, classified as general stars, young stellar objects, or cataclysmic variables. This leaves 162 objects of unknown nature, primarily clustering in the lower-left region of the near-IR color-color diagram (Fig.~\ref{fig:WR_NIR_colors_diagram}). These objects are generally faint, with approximately 80\% having magnitudes $G > $ 16 mag, and all but five are classified as WN candidates by the ESP-ELS algorithm. Sampled XP spectra of only four of these sources are available. Three of them exhibit strange continuum shapes, but one source, Gaia DR3 4655312994920412416, shows typical spectral features of a WN star. Very close to this source, a star HD 268856 that SIMBAD identifies as Brey 11, a known Magellanic WR star, is located. Our inspection of the XP spectra of these two sources indicates that SIMBAD misidentifies the object, probably because of the incorrect coordinates reported in the LMC WR catalog of \citet{1999A&AS..137..117B}. Brey 11 corresponds instead to Gaia DR3 4655312994920412416, not HD 268856 (see their \textit{Gaia} XP spectra in Fig.~\ref{fig:HD268856_false_WR}). The latest LMC WR catalog of \citet{2018ApJ...863..181N} provides correct coordinates of Brey 11 in accordance with \citet{1970CoTol..89.....S} but incorrectly associates the star with HD 268856.

Additional \textit{Gaia} XP spectra are available in continuous form. We analyzed, with the use of {\tt GaiaXPy} package\footnote{\hyperlink{https://gaia-dpci.github.io/GaiaXPy-website/}{https://gaia-dpci.github.io/GaiaXPy-website/}}, all 137 available spectra for yet unclassified sources with non-WR near-IR colors, including the four that also have sampled XP spectra. Most of these spectra exhibit unusual continua and lack clear signatures of WR stars. Only seven sources, apart from Brey 11, could be identified as probable emission-line stars, while a few others show uncertain H$\alpha$ emission. Among these, Gaia DR3 5965760637051514496 displays typical WC features. Although disqualified due to its near-IR colors (located near the WR region boundary), it remains a strong WR candidate and warrants further investigation.

Figure~\ref{fig:Ewha_rel_err} offers further insights into the clump of rejected sources. Most of these sources have very high relative errors in their pEW H$\alpha$ measurements, suggesting unreliable XP spectra (in line with our analysis of continuous XP spectra) that may have misled the ESP-ELS algorithm. Two sources of unknown nature in SIMBAD, Gaia DR3 5545207781674551424 and Gaia DR3 4206279695997383040, stand out due to their low relative errors (<10\%) in the pEW H$\alpha$ measurements, despite being relatively faint ($G \sim 16.5$ mag). The first source shows possible H$\alpha$ emission but lacks other characteristic WR features, while the second does not have an XP spectrum available for analysis.

The bottom panels of Figs.~\ref{fig:WR_NIR_colors_diagram} and \ref{fig:Ewha_rel_err} display the positions of ESP-ELS-detected sources, color-coded by their ELS Class flag. This parameter indicates the likelihood of a source being a WR star, with lower values representing higher probabilities. The clump of SIMBAD-unknown objects with non-WR near-IR colors, small pEW H$\alpha$,  and high relative error on the measurement are assigned the lowest WR probabilities.

\begin{figure}[]
   \centering
   \includegraphics[width=\columnwidth]{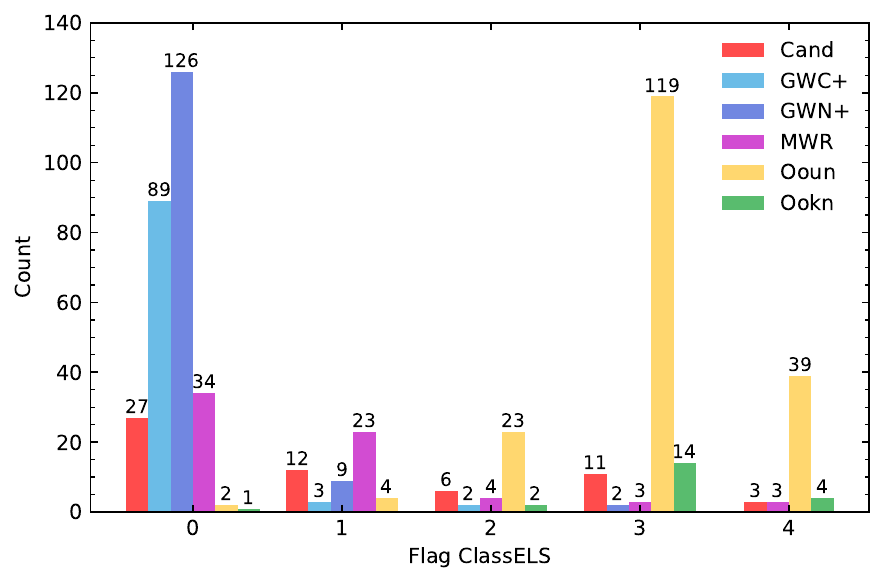}
   \caption{Number of sources in each ELS quality flag category. All 565 sources classified as WR stars by the \textit{Gaia} ESP-ELS algorithm are shown. Labels are the same as in Fig. \ref{fig:WR_NIR_colors_diagram}.}
   \label{fig:Flags_bar_chart}%
\end{figure}

The bar chart in Fig.~\ref{fig:Flags_bar_chart} illustrates the distribution of ELS Class flag categories for the 565 sources identified as WR stars by the ESP-ELS algorithm. The lowest flag categories (0 or 1) predominantly include confirmed WR stars, while the highest categories (3 or 4) contain mostly non-WR objects. However, using the ELS Class flag as a strict selection criterion at this stage would have been overly restrictive. Notably, 14 promising candidates and 8 confirmed WR stars fall into categories 3 or 4. These 22 sources are not exceptionally faint, with $G$ magnitudes ranging from 10 to 17 mag.

\subsection{Selected list of candidates}
Applying the selection criteria and filtering discussed in the previous section resulted in a list of 59 strong WR candidates, presented in Table \ref{tab:WR_Candidates_list}. These include 37 relatively bright sources with $G \leq $ 16 mag and 22 fainter candidates with $G > $ 16 mag. The magnitude division was chosen to ensure decent S/N ratios for spectral confirmation using the small-aperture telescopes available to us, guided by our prior experience with spectral confirmation of PNe \citep{2022A&A...666A.152L} and observations of symbiotic stars \citep[e.g.,][]{2022RNAAS...6...54M,2023MNRAS.523..163M,2024AN....34540017M,2023NewA...9801943P}. Observations of the fainter candidates will be conducted using larger telescopes and presented in a subsequent paper.

We note that three of our candidates, WR-C-32, WR-C-40, and WR-C-54, were recently confirmed as new WR stars by \citet{2024AJ....168..167M}. These objects remain in our list as they were identified independently by our team, and we obtained observations for two of them before their confirmation was reported in the literature.

Additionally, some of the candidates in our list have prior classifications in the literature. WR-C-01 was identified as an H$\alpha$-excess source by \citet{2021MNRAS.505.1135F}, while WR-C-02 is listed as an emission-line star \citep{1950ApJ...112...72M,1971PW&SO...1a...1S} and an eclipsing binary with a 13.6-day period \citep{2015MNRAS.452.3561L}. WR-C-34 was previously classified as an asymptotic giant branch star candidate \citep{2008AJ....136.2413R}, WR-C-36 is identified as a PN in \citep{2015MNRAS.452.2858K}, and WR-C-37 was listed as a YSO candidate by \citet{2015ApJS..220...11D}. Additionally, some targets were also categorized as long-period variables (see Table \ref{tab:WR_Candidates_list}).

\begin{table*}
    \centering
     \small 
     \setlength{\tabcolsep}{4.55pt}
     \renewcommand{\arraystretch}{1.05}
    \caption{List of selected WR stars candidates.}
    \begin{tabular}[t]{cccrrccccc}
    \hline\hline
Name    & $\alpha_{2000}$ [h:m:s]  & $\delta_{2000}$ [d:m:s] & l [°] & b [°] & \textit{G} [mag]       & pEWH$\alpha$ [nm] & SIMBAD ID  &   SIMBAD type$^{a}$     \\
    \hline
WR-C-01 & 01:40:32.96 & 63:42:22.93 & 128.339 & 1.351 & 13.87 & -2.39 & Gaia DR2 512856654641310976 & Em* \\
WR-C-02 & 06:14:34.49 & 18:28:28.27 & 192.365 & 0.533 & 11.04 & -1.10 & EM* AS 120 & s*b/EB* \\
WR-C-03 & 06:27:28.07 & 04:54:58.03 & 205.816 & -3.067 & 15.69 & -0.91 &  &  \\
WR-C-04 & 08:26:21.63 & -39:24:21.20 & 257.909 & -0.729 & 16.37 & -1.03 &  &  \\
WR-C-05 & 09:49:01.52 & -55:28:32.99 & 279.037 & -1.306 & 14.45 & -0.98 &  &  \\
WR-C-06 & 10:15:21.03 & -57:07:05.88 & 282.943 & -0.432 & 15.22 & -1.87 & IRAS 10135-5652 & LP? \\
WR-C-07 & 11:03:53.64 & -61:03:05.83 & 290.287 & -0.848 & 16.41 & -0.96 &  &  \\
WR-C-08 & 11:24:36.53 & -62:32:35.05 & 293.110 & -1.339 & 16.75 & -2.59 & 2MASS J11243656-6232350 & LP? \\
WR-C-09 & 11:54:08.07 & -62:04:48.14 & 296.279 & 0.049 & 16.28 & -0.33 &  &  \\
WR-C-10 & 11:56:20.50 & -62:34:37.81 & 296.636 & -0.381 & 14.53 & -1.24 &  &  \\
WR-C-11 & 12:56:45.51 & -64:44:45.38 & 303.500 & -1.880 & 14.63 & -1.87 & 2MASS J12564553-6444453 & LP? \\
WR-C-12 & 13:10:36.32 & -63:12:27.29 & 305.090 & -0.416 & 16.73 & -0.73 &  &  \\
WR-C-13 & 13:29:11.16 & -63:10:09.19 & 307.176 & -0.609 & 17.12 & -1.21 &  &  \\
WR-C-14 & 13:35:01.91 & -61:58:38.42 & 308.028 & 0.462 & 13.27 & -1.45 & 2MASS J13350193-6158384 & LP? \\
WR-C-15 & 14:35:08.70 & -61:14:02.54 & 315.079 & -0.832 & 16.86 & -1.99 &  &  \\
WR-C-16 & 15:16:59.19 & -59:36:37.51 & 320.404 & -1.770 & 15.98 & -1.71 &  &  \\
WR-C-17 & 15:33:17.59 & -49:48:14.22 & 327.798 & 5.090 & 16.40 & -1.10 &  &  \\
WR-C-18 & 15:37:10.77 & -56:13:18.48 & 324.521 & -0.461 & 14.75 & -1.65 & IRAS 15332-5603 & * \\
WR-C-19 & 15:37:41.30 & -57:23:35.77 & 323.886 & -1.448 & 16.23 & -1.97 &  &  \\
WR-C-20 & 15:48:50.49 & -55:11:20.18 & 326.453 & -0.635 & 16.01 & -2.17 &  &  \\
WR-C-21 & 16:48:01.37 & -42:47:38.29 & 341.993 & 1.441 & 14.42 & -1.52 &  &  \\
WR-C-22 & 16:49:41.87 & -46:45:08.71 & 339.153 & -1.332 & 16.28 & -1.69 &  &  \\
WR-C-23 & 16:59:33.96 & -47:05:36.67 & 339.948 & -2.857 & 13.45 & -0.69 & 2MASS J16593396-4705365 & LP? \\
WR-C-24 & 17:12:41.98 & -39:14:14.17 & 347.661 & -0.031 & 14.96 & -1.47 &  &  \\
WR-C-25 & 17:14:00.39 & -40:30:42.08 & 346.776 & -0.982 & 17.21 & -2.05 &  &  \\
WR-C-26 & 17:33:07.83 & -33:38:24.97 & 354.598 & -0.248 & 11.32 & -1.82 & CD-33 12168B & * \\
WR-C-27 & 17:33:23.11 & -33:43:52.46 & 354.551 & -0.342 & 16.04 & -1.38 &  &  \\
WR-C-28 & 17:33:41.28 & -33:39:48.42 & 354.642 & -0.358 & 10.83 & -0.22 & TYC 7380-64-1 & * \\
WR-C-28 & 17:35:36.42 & -33:55:57.72 & 354.632 & -0.839 & 11.60 & -0.91 &  &  \\
WR-C-30 & 17:41:27.51 & -30:16:42.56 & 358.383 & 0.065 & 16.51 & -1.20 &  &  \\
WR-C-31 & 17:59:19.86 & -21:21:59.72 & 8.094 & 1.186 & 14.59 & -4.19 &  &  \\
WR-C-32 & 18:04:52.43 & -20:37:48.65 & 9.373 & 0.424 & 12.60 & -1.94 & THA 34-2 & Em*/WR* $^b$ \\
WR-C-33 & 18:08:52.12 & -23:55:27.12 & 6.946 & -1.987 & 16.22 & -2.17 &  &  \\
WR-C-34 & 18:09:48.33 & -19:18:00.68 & 11.099 & 0.059 & 16.66 & -1.58 & 2MASS J18094833-1918006 & AB? \\
WR-C-35 & 18:14:12.97 & -19:46:02.03 & 11.189 & -1.076 & 13.73 & -0.68 & 2MASS J18141297-1946019 & LP? \\
WR-C-36 & 18:21:02.93 & -12:27:45.65 & 18.398 & 0.945 & 14.44 & -2.97 & UCAC3 156-201972 & PN \\
WR-C-37 & 18:27:10.29 & -02:57:19.80 & 27.524 & 4.038 & 15.23 & -2.19 & 2MASS J18271030-0257196 & Y*? \\
WR-C-38 & 18:28:06.09 & -10:05:01.46 & 21.309 & 0.529 & 14.66 & -1.41 &  &  \\
WR-C-39 & 18:37:07.44 & -04:25:48.79 & 27.357 & 1.155 & 14.87 & -2.18 &  &  \\
WR-C-40 & 18:40:14.97 & -05:03:20.12 & 27.159 & 0.176 & 12.75 & -0.69 & THA 14-54 & Em*/WR* $^b$ \\
WR-C-41 & 18:40:35.23 & -09:41:39.52 & 23.072 & -2.021 & 15.75 & -1.35 &  &  \\
WR-C-42 & 18:43:22.18 & -01:03:37.30 & 31.068 & 1.311 & 15.11 & -1.70 &  &  \\
WR-C-43 & 18:45:17.38 & -02:48:13.10 & 29.736 & 0.088 & 14.97 & -2.19 &  &  \\
WR-C-44 & 18:46:15.05 & -05:42:21.13 & 27.263 & -1.449 & 15.42 & -1.73 &  &  \\
WR-C-45 & 18:53:05.40 & -02:08:33.18 & 31.213 & -1.344 & 17.20 & -1.16 &  &  \\
WR-C-46 & 18:53:19.55 & 03:32:52.44 & 36.305 & 1.198 & 13.68 & -0.56 &  &  \\
WR-C-47 & 19:02:21.70 & 08:47:09.20 & 41.992 & 1.590 & 15.97 & -1.19 &  &  \\
WR-C-48 & 19:31:15.07 & 19:04:18.30 & 54.358 & 0.207 & 15.28 & -1.17 &  &  \\
WR-C-49 & 19:32:13.42 & -54:05:07.01 & 343.537 & -27.498 & 17.58 & -0.49 &  &  \\
WR-C-50 & 19:34:10.42 & 19:22:13.94 & 54.953 & -0.253 & 13.93 & -0.92 & 2MASS J19341042+1922141 & LP? \\
WR-C-51 & 19:38:34.47 & 24:03:20.48 & 59.541 & 1.140 & 17.60 & -1.90 & 2MASS J19383447+2403205 & LP? \\
WR-C-52 & 19:52:40.81 & 23:29:33.90 & 60.672 & -1.934 & 16.47 & -1.76 & 2MASS J19524081+2329339 & LP* \\
WR-C-53 & 20:09:08.72 & 36:38:00.92 & 73.711 & 1.980 & 16.76 & -4.61 & ZTF J200908.72+363801.0 & LP* \\
WR-C-54 & 20:43:36.48 & 44:55:05.16 & 84.236 & 1.472 & 10.34 & -0.46 & LS III +44 21 & EB*/WR* $^b$ \\
WR-C-55 & 20:57:40.95 & 46:59:44.99 & 87.394 & 0.902 & 15.44 & -2.44 &  &  \\
WR-C-56 & 21:05:28.26 & 51:36:00.76 & 91.692 & 2.990 & 17.23 & -2.36 & ZTF J210528.26+513600.7 & LP* \\
WR-C-57 & 21:19:47.42 & 53:22:25.00 & 94.481 & 2.622 & 15.76 & -3.06 &  &  \\
WR-C-58 & 21:27:21.23 & 52:32:05.06 & 94.705 & 1.226 & 15.20 & -0.70 &  &  \\
WR-C-59 & 22:09:50.11 & 60:22:33.67 & 104.210 & 3.518 & 15.20 & -0.92 &  &  \\
\hline
    \end{tabular}
    \tablefoot{
            $^{a}$SIMBAD types correspond to the following classes: Em* - emission-line star; s*b - blue supergiant; EB* - eclipsing binary; LP*/LP? - long-period variable/candidate; * - star; WR* - WR star; AB? - asymptotic giant branch star candidate; PN - planetary nebula; Y*? - young stellar object candidate. $^{b}$The type was updated to WR* after the first submission of this paper based on the classification by \citet{2024AJ....168..167M}; see the text for details.}
    
    \label{tab:WR_Candidates_list}
\end{table*}

\begin{figure}[]
    \centering
        \includegraphics[width=0.7\columnwidth]{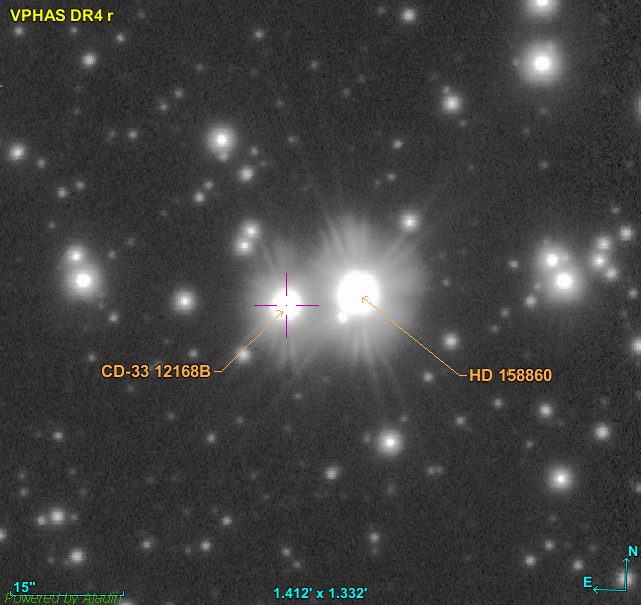}
    \caption{Field of WR94 from VPHAS+ survey. Marked are HD 158860 and CD-33 12168B (correct counterpart of WR94), discussed in the text.}
    \label{fig:WR94}
\end{figure}

 %--------------------------------------------------------------------
\section{Spectroscopic follow-up}
\label{sec:Spectroscopic follow-up}
We conducted spectroscopic observations of all bright candidates ($G \leq $ 16 mag) using facilities of the Southern Spectroscopic Project Observatory Team (2SPOT) consortium and through access to the Calern Observatory, affiliated with the Observatoire de la Côte d’Azur (OCA). Observations were carried out from August to December 2024 (see Table \ref{tab:Observed_WR_Candidates}).

The 2SPOT consortium\footnote{\hyperlink{http://www.2spot.org}{http://www.2spot.org}} was established in 2019 by five French amateur astronomers and operates two remotely controlled telescopes in Chile. These include a 0.3-m Ritchey-Chrétien telescope equipped with a medium-resolution Eshel spectrograph (R = 10\,000) and a 0.3-m F/4 Newtonian telescope equipped with a low-resolution Alpy600 spectrograph (R = 600). For this study, we utilized the Alpy600 spectrograph with a 23 $\mu$m slit. The telescopes are located at Deep Sky Chile (DSC), near Cerro Tololo Observatory.

For northern hemisphere candidates, we used our observatories in Kermerrien and Cornillon, France. These setups are similar to the Chile-based ones but use smaller 0.2 m F/5 Newtonian telescopes. Observations were recorded with ATIK 414 EX cooled cameras with 1392 × 1040 pixel sensors with a pixel size of 6.45 $\mu$m. This configuration yields a dispersion of approximately $\sim$3.0 \AA/pixel,\,and covers a spectral range of 3800 to 7800~\AA.

We were also granted access to the C2PU's 1-m F/6.8 Cassegrain Epsilon telescope at the Calern Observatory. Observations were conducted using a LISA spectrograph with a 50 $\mu$m slit (R = 500) and an ATIK 414 EX cooled camera. This configuration yields a dispersion of $\sim$2.5 \AA/pixel and a spectral range of 4000 to 7500 \AA.

Observations typically used subframes of 1200 s at bin 1$\times$1, with total exposure times ranging from 40 minutes to several hours, depending on the magnitude of the target. Data reduction was performed using the {\tt Integrated Spectrographic Innovative Software} (ISIS)\footnote{\hyperlink{http://www.astrosurf.com/buil/isis-software.html}{http://www.astrosurf.com/buil/isis-software.html}}. All spectra were corrected for dark frames, bias, and flat fields using standard techniques. The instrumental response was determined using reference spectra of flux standard stars obtained under the same observing conditions as the targets. Wavelength calibration was performed using an argon-neon lamp. No telluric correction was applied to the final spectra.

%--------------------------------------------------------------------
\section{Results and spectral classification}
\label{sec:Results and spectral classification}
\begin{figure}[]
    \centering
        
        \includegraphics[width=\columnwidth]{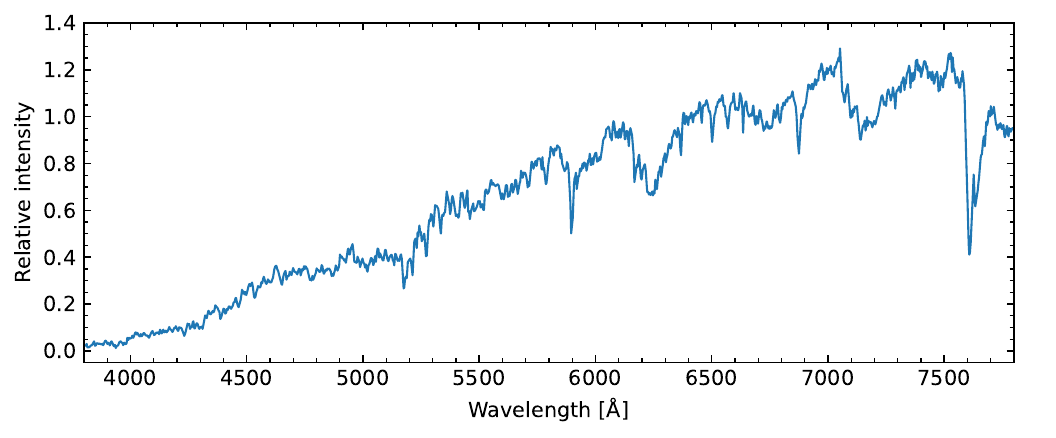}
    \caption{Our spectrum of HD 158860, incorrectly associated with WR94 in SIMBAD, in the GWRC, and subsequently in several published works.}
    \label{fig:WR94_spec}
\end{figure}

We observed\footnote{All the spectra are available online at: \hyperlink{https://2spot.org/EN/WR\_spectra\_list.php}{https://2spot.org/EN/WR\_spectra\_list.php} and will be available also at CDS.} all 37 of our bright candidates ($G \leq $ 16 mag), confirming the majority as WR stars. Of these, 33 are newly identified WR stars, while one source, WR-C-26, was already known as WR94. However, we uncovered a long-standing misidentification of WR94 in the literature: the GWRC and SIMBAD incorrectly associate WR94 \citep[also referred to as Hen 3-1420;][]{1976ApJS...30..491H} with HD 158860 (=CD-33 12168). Consequently, several statistical studies included the incorrect counterpart of WR94. Our observations confirm that WR94 is actually CD-33 12168B, a star located less than 10 arcseconds from HD 158860 (see the field in Fig. \ref{fig:WR94}), in line with the original report of \citet{1976ApJS...30..491H}. The confusion likely stems from table VII in \citet{1981SSRv...28..227V}, where the 'B' in the identifier CD-33 12168B was omitted, despite the correct position being shown in their finding chart. We positioned both stars within the slit of our spectrograph and confirmed unequivocally that WR94 corresponds to CD-33 12168B. The spectra of HD 158860 and CD-33 12168B (WR94) are shown in Figs. \ref{fig:WR94_spec} and \ref{fig:WN_classified} (see WR-C-26).

Our main results are summarized in Table \ref{tab:Observed_WR_Candidates} and discussed in detail in subsequent sections.
We spectroscopically confirmed 17 WC stars and 16 WN stars, with subclasses in the WC5–9 and WN3–8 ranges, respectively. The observed spectral classifications align with the WC or WN ELS classes assigned by the ESP-ELS algorithm.

Our diagnostic relied on identifying and analyzing key emission lines within the spectra and on the relative strength of those close enough in wavelength to be compared qualitatively (e.g. \ion{C}{iii} $\lambda$5696 and \ion{C}{iv} $\lambda$5806 for WC type, or \ion{N}{iii} $\lambda$4640 and \ion{He}{ii} $\lambda$4686 lines for WN type). We primarily followed the classification scheme presented in table 2 of \citet{2001NewAR..45..135V}, as in a large fraction of the spectra, the fainter lines were too difficult to measure to apply the quantitative schemes of \citet{1990ApJ...358..229S} or \citet{1998MNRAS.296..367C} for WC stars, and \citet{1996MNRAS.281..163S} for WN stars.

\subsection{Newly confirmed WC stars}

The classification of WC stars \citep{1987ApJS...65..459T, 1998MNRAS.296..367C, 2001NewAR..45..135V} relies on the relative strength of the two closely spaced \ion{C}{iii} $\lambda$5696 and \ion{C}{iv} $\lambda$5806 lines, which are less affected by dust extinction. The \ion{C}{iii} $\lambda$5696 line tends to be stronger than \ion{C}{iv} $\lambda$5806 in late subclasses, while \ion{C}{iv} $\lambda$5806 dominates in early subclasses. The full width at half maximum (FWHM) of these lines increases from late to early WC types. The \ion{C}{ii} $\lambda$4267 line is also useful for distinguishing between late subclasses WC8 and WC9. Unfortunately, this line is strongly affected by dust extinction and is undetectable or very weak in our data.

The spectra plotted in Fig. \ref{fig:WCL} exhibit very similar characteristics. The \ion{C}{iii} $\lambda$5696 emission is relatively strong compared to \ion{C}{iv} $\lambda$5806. The \ion{C}{ii} $\lambda$4267 line is not detected in any spectrum, likely due to dust extinction as the \ion{C}{ii} $\lambda$7231 line appears very strong in the red part of each spectrum. The intensity of \ion{C}{iii} $\lambda$4650 also suffers from high reddening. \ion{O}{v} $\lambda$5592 is either absent or very weak. For most spectra, the FWHM of the \ion{C}{iii} $\lambda$5696 line ranges from 17 to 30 \AA, except for WR-C-31, which has a larger FWHM of 60 \AA. The flat-topped profile of this line in WR-C-31 indicates that this star belongs to the WC8 subclass \citep{1987ApJS...65..459T}.WR-C-46 and WR-C-16 also potentially belong to the WC8 subclass according to their \ion{C}{iv} $\lambda$5806 / \ion{C}{iii} $\lambda$5696 ratio. All other stars belong to the late WC9 subclass, according to the classification scheme discussed above. WR-C-37 (WC9) was previously identified as a YSO candidate in \cite{2015ApJS..220...11D}, but our spectrum does not support this conclusion.

The spectra of the stars shown in Fig. \ref{fig:WC5_7} are dominated by \ion{C}{iv} $\lambda$5806 emission, with \ion{C}{iii} $\lambda$5696 line being stronger or comparable to \ion{O}{v} $\lambda$5592. These spectral features rule out very late and early subclasses such as WC8, WC9, or WC4. WR-C-05, WR-C-10, and WR-C-58 exhibit very strong \ion{C}{iv} $\lambda$5806 emission compared to \ion{C}{iii} $\lambda$5696, and their \ion{O}{v} $\lambda$5592 line is comparable to or even slightly stronger than \ion{C}{iii} $\lambda$5696 when corrected for dust extinction. For these reasons, we classified these stars as WC5. The multiple absorption lines in the WR-C-10 spectrum suggest the possible presence of a companion. WR-C-47 likely belongs to the WC6 subclass, as the prominence of \ion{C}{iv} $\lambda$5806 is less marked than in the previously discussed spectra, and its \ion{C}{iii} $\lambda$5696 emission is more intense than \ion{O}{v} $\lambda$5592. The other stars, with \ion{C}{iii} emission much stronger than \ion{O}{v}, belong to the WC7 subclass.

\begin{figure}[]
   \centering
   \includegraphics[width=0.49\textwidth]{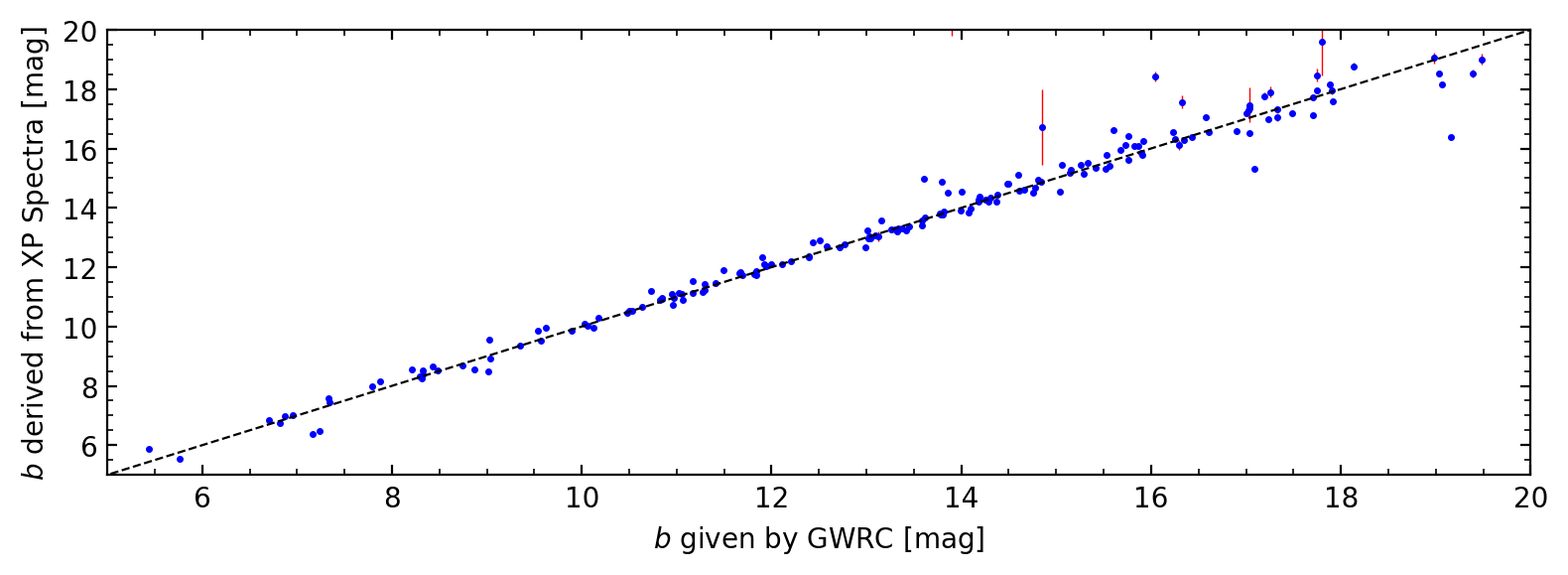}
   \hspace{0.005\textwidth}
   \includegraphics[width=0.49\textwidth]{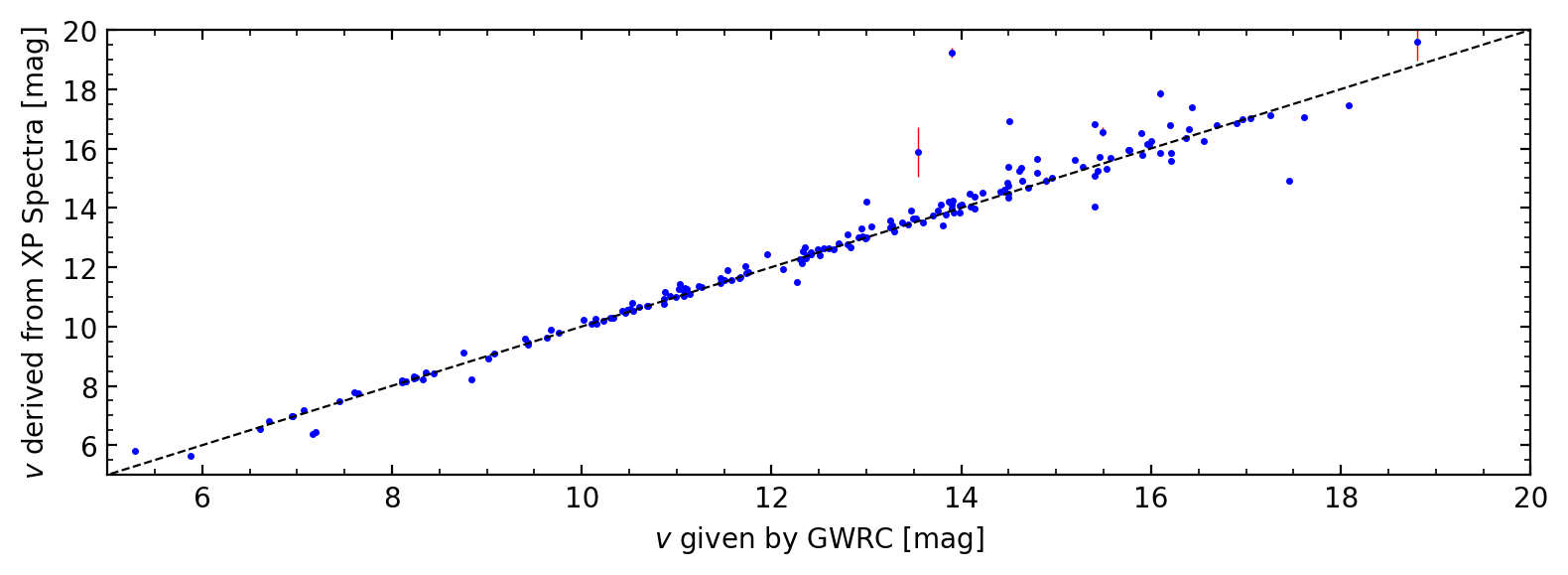}
   \caption{Comparison of \textit{b} and \textit{v} magnitudes provided by the GWRC and derived from \textit{Gaia} XP spectra.}
   \label{fig:bv_mag_comparison}%
\end{figure}

\begin{figure*}[]
    \centering
        \includegraphics[width=0.8\textwidth]{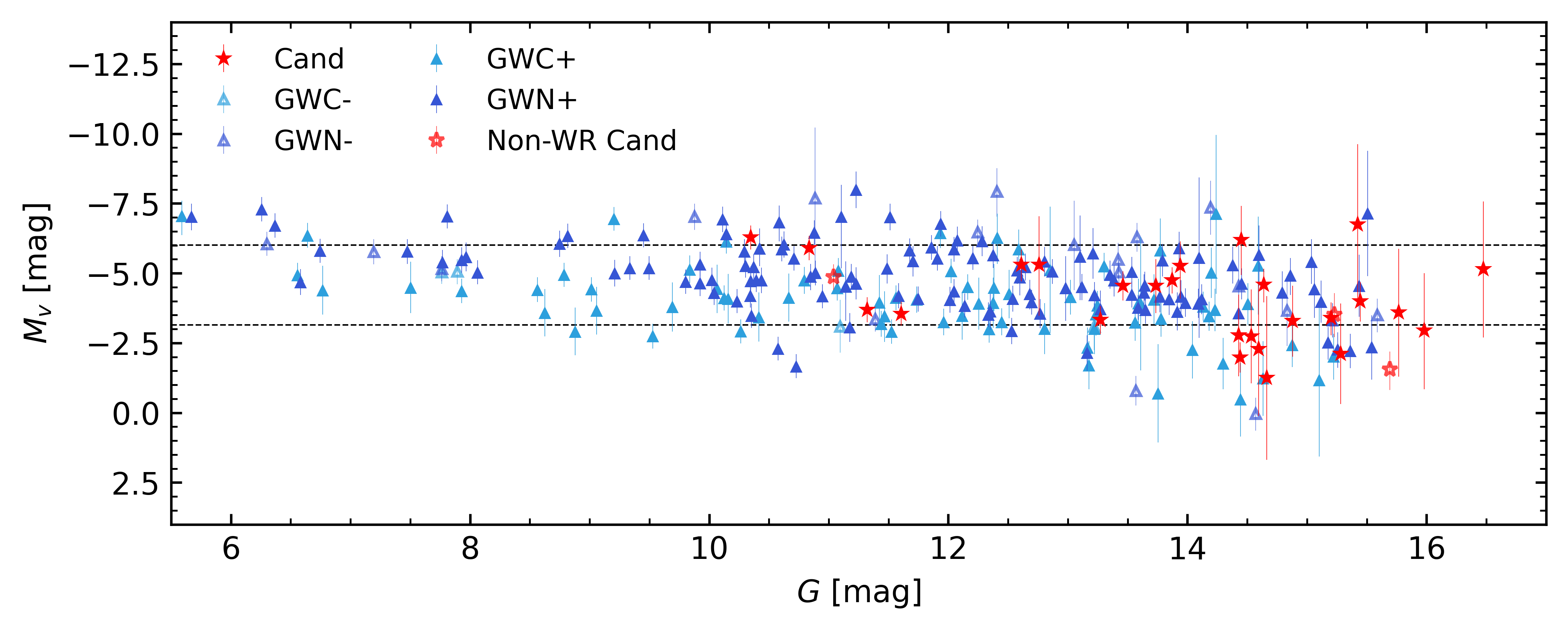}
    \caption{Absolute magnitude \(M_{v}\) of known WR stars and our candidates (see Fig. \ref{fig:WR_NIR_colors_diagram} for the labels legend. The candidates that we confirmed as non-WR are marked with empty red stars symbols). Magnitudes are derived from the \textit{Gaia} XP spectra. The black dashed lines denote the lower and upper limits of the mean value we derived for known Galactic WR stars (i.e., the sample made of GWC+, GWN+, GWC-, and GWN-).}
    \label{fig:Abs_Mag_v_all}
\end{figure*}

\subsection{Newly confirmed WN stars}

The classification of WN stars is primarily based on the comparison of the nitrogen lines of \ion{N}{iv} $\lambda$4057, \ion{N}{v} $\lambda$4604, and \ion{N}{iii} $\lambda$4640 \citep{1995A&AS..113..459H, 1996MNRAS.281..163S, 2001NewAR..45..135V} in the blue part of the spectrum. Our candidates suffer from high interstellar reddening, which often makes the detection of \ion{N}{v} or \ion{N}{iv} lines uncertain, thereby complicating the classification. Note that \ion{He}{ii} $\lambda$5411 and \ion{C}{iv} $\lambda$5806 become stronger from late WN7 to early WN4 subclasses, while \ion{He}{i} $\lambda$5875 tends to be very weak or completely absent in subclasses earlier than WN4. The spectra of WN stars for which a subclass can be determined are shown in Fig. \ref{fig:WN_classified}. 

WR-C-01 likely belongs to the WN4 subclass since our spectrum shows \ion{N}{v} $\lambda$4604 and no \ion{N}{iii} $\lambda$4640. WR-C-11 shows a \ion{He}{ii} $\lambda$4686 line stronger than \ion{N}{iii} $\lambda$4640 and a relatively intense \ion{He}{ii} $\lambda$5411. WR-C-11 likely belongs to the WN5 subclass. WR-C-42 shares similar characteristics but with a stronger \ion{He}{i} $\lambda$5875, indicating that this star rather belongs to the WN6 subclass.

WR-C-26 corresponds to WR94 (cf. Section \ref{sec:Results and spectral classification}), a known WN5o star according to the GWRC (v1.29). However, we classify this star as a potential WN6 since our spectrum shows the presence of a faint \ion{N}{v} $\lambda$4604, remaining much weaker than the \ion{N}{iii} $\lambda$4640.

WR-C-54 (LS III +44 21) is a recently confirmed WN6+O6.5 binary system \citep{2024AJ....168..167M}. The spectrum of WR-C-28 is very similar to WR-C-54, showing \ion{He}{ii} $\lambda$4686 stronger than \ion{N}{iii} $\lambda$4640, absence of \ion{C}{iv} $\lambda$5806 and \ion{He}{i} $\lambda$5875, and many absorption lines. WR-C-28 may also be a binary system with a WN6 companion.

The spectrum of WR-C-55 is marked by the strong and broad emission line of \ion{He}{ii} $\lambda$4686, a very weak \ion{N}{v} $\lambda$4604 emission and the absence of \ion{N}{iii} $\lambda$4640 and \ion{He}{i} $\lambda$5875. WR-C-55 belongs to the WN3 subclass. The spectrum of WR-C-57 shares similar characteristics but with faint \ion{N}{iii} $\lambda$4640 and \ion{He}{i} $\lambda$5875 in emission and an unusually strong \ion{C}{iv} $\lambda$5806 line. This indicates that WR-C-57 is a transition-type WN/C star (see \citeauthor{1989SPIE.1140..376C} \citeyear{1989SPIE.1140..376C} for more information on WN/C class).

The remaining stars most likely belong to late WN subclasses. WR-C-29 and WR-C-36 have \ion{He}{ii} $\lambda$4686 slightly stronger than \ion{N}{iii} $\lambda$4640 and weak \ion{He}{i} $\lambda$5875, suggesting they belong to the WN7 subclass. The H$\beta$ line of WR-C-36 is stronger than \ion{He}{ii} $\lambda$5411, indicating the star is hydrogen-rich. \cite{2015MNRAS.452.2858K} previously considered a WN7h classification for this star based on near-IR observations (referred to as 1430-AB0 in their study), but they classified this source as a PN at the end due to the absence of P Cygni absorption lines typically associated with WN7h stars. However, we do not support this conclusion, as WR-C-36 is too luminous for being a PN (although slightly fainter than expected for the WN7h star; cf. Sect. \ref{subsec:Absolute magnitude} and \ref{subsec:Absolute_infrared}, and Tables \ref{tab:absolute_mags} and \ref{tab:absolute_mags_IR}) and also shows no evidence of PN nebular emission lines in its spectrum (Fig. \ref{fig:WN_classified}) and PN features in the mid-IR domain (cf. Section \ref{subsec:IR properties}). Furthermore, the object is not included in the HASH PN Database \citep{2017IAUS..323..327B}.

The \ion{He}{ii} $\lambda$4686 and \ion{N}{iii} $\lambda$4640 lines of WR-C-23 and WR-C-35 have comparable intensities. Their \ion{He}{i} $\lambda$5875 line is also relatively strong, suggesting that these stars belong to the WN8 subclass.

The spectra shown in Fig. \ref{fig:WN_likely} have a S/N ratio too low to precisely determine the subclasses of the stars, but their spectral features strongly suggest they are WN stars. WR-C-59 is likely a weak-lined WN3 star, with a possible detection of \ion{N}{v} $\lambda$4604. WR-C-44 and WR-C-50 exhibit weak \ion{He}{ii} $\lambda$4686 emission and no \ion{N}{iii} $\lambda$4640, suggesting these stars are likely in the WN5-6 subclass range. It is also likely the case of WR-C-41, since \ion{N}{iii} $\lambda$4640 appears stronger than \ion{N}{v} $\lambda$4604. In the spectrum of WR-C-39, \ion{N}{iii} $\lambda$4640 is detected but remains slightly weaker than \ion{He}{ii} $\lambda$4686, indicating that this star may belong to the WN7-8 subclass range.

\subsection{Objects of non-WR nature}

The spectra of three objects in our sample, which do not exhibit characteristics of WR stars, are grouped in Fig. \ref{fig:uncertain_nature}. These spectra display narrow \ion{H}{i} and \ion{He}{i} emission lines (limited to H$\alpha$ in the case of WR-C-03) against a flat continuum. Notably, WR-C-02 is a previously identified eclipsing binary system, where the emission lines are distinctly marked by a prominent P Cygni profile.

 %--------------------------------------------------------------------
\section{Distinguishing between WR and [WR] stars}
\label{sec:Population I massive WR stars or Population II low mass [WR]}
Here, we analyze the absolute magnitudes, vertical $z$ distances from the Galactic mid-plane, and infrared properties of the newly confirmed WR stars and candidates to distinguish massive Population I WR stars from Population II low-mass [WR] stars, which may exhibit similar spectroscopic features.

\subsection{Absolute \(M_{v}\) magnitudes  of newly confirmed WR stars and candidates}
\label{subsec:Absolute magnitude}

The assessment of absolute magnitudes of WR stars depends on parameters like reddening and distance, which are not always precisely known. Without accurate values for these factors, the analysis can suffer from significant uncertainties. Our aim was to estimate the absolute magnitudes of our candidates in the 
\textit{v} \citep{1968MNRAS.140..409S} and $K_{s}$ bands and to compare the results with those of known WR stars.

For calculating absolute magnitudes in the \textit{v} band, we followed the methodology outlined in \cite{2020MNRAS.493.1512R}. This approach utilizes the $(b-v)$ measurement, where \textit{b} and \textit{v} are monochromatic magnitudes at 4270 and 5160 \AA, respectively. These wavelengths were chosen to minimize contamination from emission lines \citep{1968MNRAS.140..409S}, which can strongly influence reddening estimates. The $(b-v)$ colors in our analysis were derived from the XP spectra of \textit{Gaia} using the {\tt GaiaXPy} library. Notably, $(b-v)$ can also be obtained from our data even when the spectra are not calibrated in absolute fluxes. We compare the results at the end of the section.

To evaluate the reliability of magnitudes derived from \textit{Gaia} XP spectra, we compared the \textit{b} and \textit{v} magnitudes of known WR stars, computed from XP spectra\footnote{The WR filters were approximated using filters with similar properties from the \textit{Spanish Virtual Observatory} filter profile service.}, with magnitudes provided by the GWRC based on ground-based observations. Figure~\ref{fig:bv_mag_comparison} includes data for approximately 180 known WR stars, for which \textit{b} and/or \textit{v} magnitudes were available both in the GWRC and derivable from XP spectra. The comparison shows excellent consistency between the datasets, supporting the use of XP spectra for assessing $M_{v}$ with reasonable confidence.

The color excess \(E(b-v)\) was calculated using typical intrinsic \((b-v)_{0}\) colors for WR stars, as listed in table 4 of \cite{2020MNRAS.493.1512R}. These intrinsic colors depend on the WR subclass and binarity. For single stars, \((b-v)_{0}\) ranges from $-0.18$ to $-0.37$ mag. For bright sample candidates with observed spectra, we assigned \((b-v)_{0}\) based on the subclasses determined in Sect. \ref{sec:Results and spectral classification}, assuming they are single WR stars. For faint sample candidates not yet observed (both WC and WN according to ELS classification), we adopted a representative average value of \((b-v)_{0}=-0.3\pm0.1\) mag. For the three candidates confirmed as non-WR stars, we arbitrarily adopted the same average value. The extinction $A_{v}$ was computed assuming a standard \(R_{v}=4.12\) (see \cite{1982IAUS...99...57T}). The resulting \(M_{v}\) was calculated using distances from the \citet{2021AJ....161..147B} catalog (see Table \ref{tab:paral}).

Uncertainties in \(M_{v}\) include errors in the \(b, v\) fluxes, intrinsic colors \((b-v)_{0}\), and distances. Flux uncertainties in $b$ and $v$ were derived from the \textit{Gaia} XP spectra, while uncertainties in 
\((b-v)_{0}\) were taken from table 4 of \cite{2020MNRAS.493.1512R}. For distances, we used the 16th and 84th percentiles of the geometric distance estimates from \cite{2021AJ....161..147B}.

The absolute magnitudes \(M_{v}\) of the candidates are summarized in Table \ref{tab:absolute_mags}. For candidates with unreliable \textit{Gaia} data or missing XP spectra (both sampled and continuous forms), we did not compute magnitudes. As a result, these 20 targets have no entries in Table \ref{tab:absolute_mags}. Additionally, no distance measurement is available for WR-C-49, so \(M_{v}\) could not be determined\footnote{The galactic latitude of the source (b = -27.5\textdegree; see Table \ref{tab:WR_Candidates_list}) suggests that this star is a rather poor WR candidate.}. Figure~\ref{fig:Abs_Mag_v_all} compares the absolute magnitudes \(M_{v}\) of known WR stars and our candidates. 
In the figure, we excluded 9 candidates with low S/N XP spectra, which resulted in \(M_{v}\) uncertainties exceeding 4 mag (see Table \ref{tab:absolute_mags}).

The mean values \(\overline{M}_{v}\) obtained for known Galactic WR stars was found to be \(-4.6 \pm 1.4\) mag, with sample standard deviation as uncertainty. Population II [WR] stars are significantly fainter, given their luminosities are at least an order of magnitude lower than those of Population I WR stars. According to table 5 of \cite{2020MNRAS.493.1512R}, \(\overline{M}_{v}\) for WN stars ranges from $-3.6$ to $-7.0$ mag, and for WC stars, it ranges from $-3.9$ to $-4.6$ mag. Table 4 of \cite{2020MNRAS.493.1512R} gives a mean value of \(-4.3 \pm 1.8\) mag, consistent with our results obtained using \textit{Gaia} XP spectra.

Most candidates plotted within the range (see Fig. \ref{fig:Abs_Mag_v_all}) likely belong to Population I WR stars. All candidates with \(M_{V} > -3\) mag and \(G > 14\) mag were spectroscopically observed, and all but one exhibit clear WR spectral features. WR-C-03, with \(G = 15.69\) mag and \(M_{v}=-1.6\pm0.7\) mag, significantly deviates from the WR sample (and the uncertainty of its absolute magnitude is low). Our observations confirm that WR-C-03 is not a WR star (cf. Fig.~\ref{fig:uncertain_nature}).

Table \ref{tab:absolute_mags} also lists absolute magnitudes calculated using $(b-v)$ colors derived from our spectra. For stars with $G<14$ mag, the results agree well with those obtained solely from XP spectra. For fainter stars, the low S/N in the blue part of our spectra precludes this analysis.

\subsection{Absolute \(M_{K_{s}}\) magnitudes}
\label{subsec:Absolute_infrared}

In addition to $M_v$, we also evaluated $M_{K_{s}}$, following the approach described by \citet{2015MNRAS.447.2322R} and \citet{2020MNRAS.493.1512R}. The apparent near-IR magnitudes are taken from the 2MASS catalog. From these data and intrinsic colors of WR stars of various subtypes \citep[see table 4 of][]{2015MNRAS.447.2322R}, we calculated color excesses \(E(J-K_s)\) and \(E(H-K_s)\). If the subtype of a particular candidate is known from our observations (Sect. \ref{sec:Results and spectral classification}), we adopted the intrinsic color typical for that subtype. For the faint, yet unobserved candidates (and for three non-WR stars), we used the average color of the WC or WN stars assigned based on the ELS class of the candidate. If that was the case, it is indicated in Table \ref{tab:absolute_mags_IR}.
 
From the color excesses, we calculated two values of $K_s$ band extinction using equations 1 and 2 of \citet{2015MNRAS.447.2322R}, in the same way as presented in the aforementioned paper. Finally, we calculated $M_{K_s}$ magnitudes by adopting distances from Table \ref{tab:paral}. The two extinction values were averaged for the calculation. The uncertainties in the absolute magnitudes include the uncertainties in the input 2MASS magnitudes, in the extinction laws, in the distances, and in the average extinction value adopted.

We were able to calculate $M_{K_s}$ for all candidates except WR-C-49, which does not have any distance estimate available. The magnitudes for all new WR stars classified in this work (cf. Table \ref{tab:absolute_mags_IR}) are consistent with those of known WR stars, as shown in table 5 and figs. 1 and 2 of \citet{2015MNRAS.447.2322R}. Magnitudes of WR-C-18, WR-C-24, WR-C-32, and WR-C-43, in the range of $-7.8$ to $-8.8$ mag, indicate that these might belong to WC9d types (with additional circumstellar dust).

From the faint sample, 9 targets classified as WC stars by \textit{Gaia} ELS show magnitudes within the typical range of WC stars. The magnitude of WR-C-30 suggests the presence of circumstellar dust. From 13 objects with WN ELS class, several fall outside the typical range of $\sim$($-2.5$ to $-7.0$) mag \citep{2015MNRAS.447.2322R}, including WR-C-4, WR-C-9, WR-C-12, WR-C-17, and WR-C-53. None of these targets have $M_v$ magnitudes listed in Table \ref{tab:absolute_mags}. The XP spectra for these objects are not sufficiently conclusive to rule out the potential WR nature of these candidates. Consequently, we defer further discussion of the faint sample to a subsequent paper, where we will present spectroscopic follow-up observations of these sources.

\subsection{Vertical distances from the Galactic disc}
Young massive stars are typically confined to the Galactic thin disc \citep[see, e.g., fig. 6 in][]{2015MNRAS.447.2322R}, whereas significantly older CSPNe are often found at larger vertical distances from the disc. We have, therefore, calculated the $z$ distances for all candidates using the {\tt astropy} package and adopting distances from \citet{2021AJ....161..147B}, except for WR-C-49, for which no distance is available. The results are summarized in Table \ref{tab:paral}.

Among the newly confirmed WR stars, the majority lie within 200 pc of $z = 0$, though several are located at larger distances: WR-C-05 at $-270^{+70}_{-81}$ pc, WR-C-11 at $-338^{+51}_{-69}$ pc, WR-C-16 at $-201^{+46}_{-64}$ pc, WR-C-37 at $402^{+97}_{-79}$ pc, WR-C-41 at $-290^{+57}_{-84}$ pc, WR-C-57 at $281^{+55}_{-40}$ pc, and WR-C-59 at $533^{+103}_{-83}$ pc. WR-C-59, in particular, might belong to the WR stars with the largest $z$ distances. However, caution is necessary when interpreting these results, as most of these stars have low $\varpi$/$\sigma_\varpi$ $\approx$ 1--4 (S/N of the parallax), making their distance estimates potentially unreliable. Notably, the parallax of WR-C-11 in \textit{Gaia} DR3 is negative.

Several faint, unobserved candidates also have $|z| > 200$ pc. Noteworthy is WR-C-17, with $z = 881^{+228}_{-187}$ pc. This object was noted in the previous section as having an $M{K_S}$ value fainter than is typical for WR stars.

\subsection{Mid-infrared properties}
\label{subsec:IR properties}

\begin{figure}[]
    \centering
        \includegraphics[width=0.47\columnwidth]{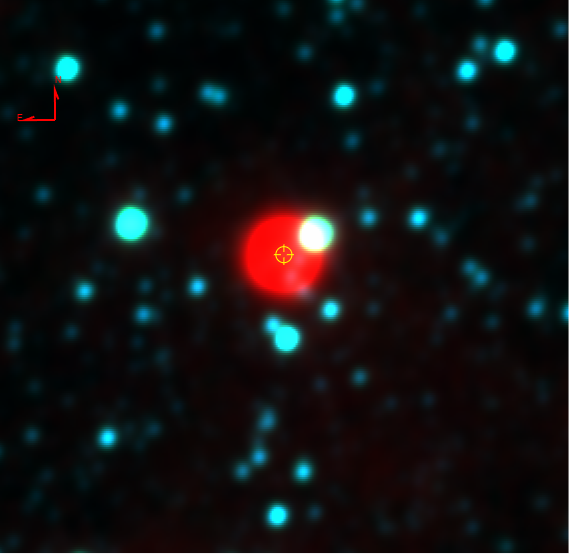}
    \hspace{0.005\textwidth}
        \includegraphics[width=0.47\columnwidth]{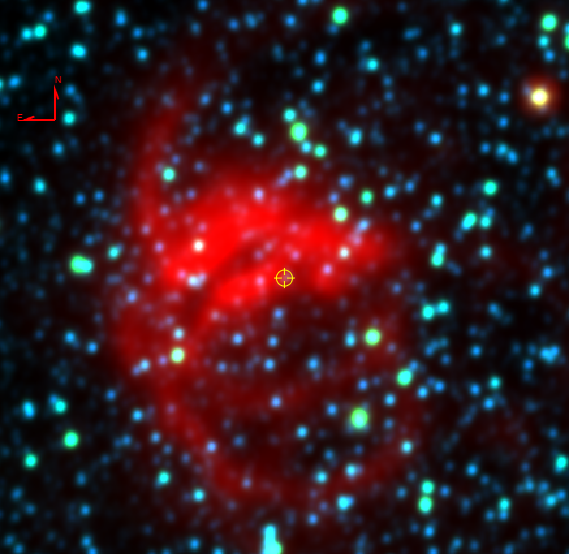}
    \caption{Typical morphology of WR and [WR] nebulosities. The images were constructed on the basis of WISE observations with the red, green, and blue channels corresponding to \textit{W4} (24 µm), \textit{W2} (4.6 µm), \textit{W1} (3.4 µm) filters, respectively. \textbf{Left panel:} RaMul2 [WR] PN whose colors and morphology are typical of a young and compact PN. The field of view is 6' $\times$ 5'. \textbf{Right panel:} WR102 ring nebula showing a complex morphology made of several bubbles. The field of view is 10' $\times$ 10'.}
    \label{fig:PN_vs_WR_ring}
\end{figure}

The properties of [WR] PNe in the mid-infrared provide useful observational information that can further support the classification of Population I or II WR stars. PNe are characterized by intense emission in the mid-IR region \citep{2010PASA...27..129F}, generated by the dust ejected by the PN progenitor and heated by its remnant core. This emission is particularly strong for compact and young PNe, and it is easily observable as a compact, perfectly circular-shaped emission in the Wide-field Infrared Survey Explorer \citep[WISE;][]{2010AJ....140.1868W} \textit{W4} band (24 µm) (cf. Fig.~\ref{fig:PN_vs_WR_ring}). This is a key observational feature often employed in the search for new PNe \citep[e.g.,][]{2022A&A...666A.152L}.

We checked each PN associated with a [WR] central star in the catalog of \cite{2020A&A...640A..10W}. All of them show the aforementioned typical mid-IR emission. This is not surprising, as the evolutionary scheme of a [WC] central star is assumed to go from late [WC], through early [WC] and PG1159 stars, to a white dwarf \citep{2000A&A...364..597G}. Given that [WR] PNe are at the beginning of their PN phase, they are compact and dense.

Some Population I WR stars are associated with a ring nebula made of ejected material. Only a few massive WR stars exhibit this feature. WR ring nebulae also emit in mid-IR but produce various kinds of bubbles, arcs, clumps, or amorphous shapes. Three broad morphological types are defined in \cite{2015A&A...578A..66T}. Their mid-IR observational properties are very different from those of compact PNe (cf. Fig.~\ref{fig:PN_vs_WR_ring}), but they can be confused with evolved or well-resolved PNe.

The WISE colors of our candidates do not correspond to those of compact PNe, and none of them show prominent emission in the WISE \textit{W4} band. This supports the conclusion that our candidates are not [WR] CSPNe and most likely belong to massive Population I WR stars.

\begin{figure}[]
    \centering
    \includegraphics[width=0.47\columnwidth]{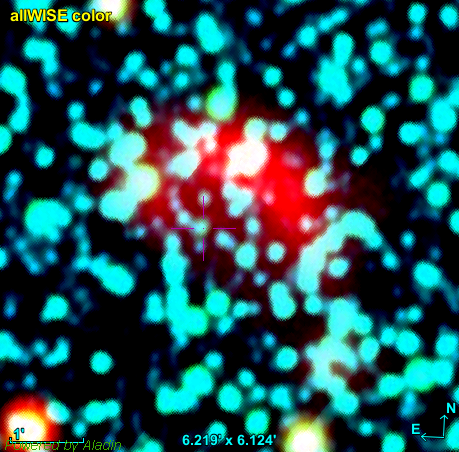}
    \hspace{0.005\textwidth}
    \includegraphics[width=0.47\columnwidth]{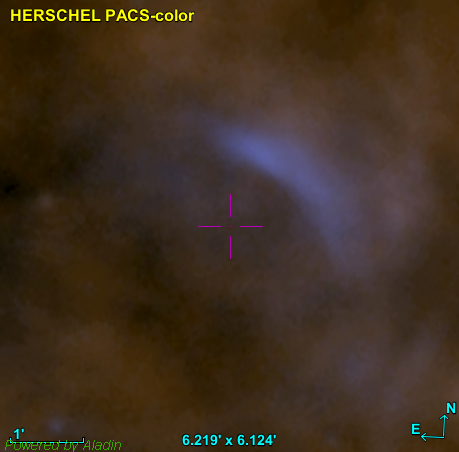}
    \caption{Arc-shaped mid-IR emission northwest of the WR-C-27. \textbf{Left panel:} WISE color image constructed in the same way as images in Fig. \ref{fig:PN_vs_WR_ring}. \textbf{Right panel:} Herschel PACS color image. The red, green, and blue channels correspond to observations at 160 µm, 100 µm, and 70 µm, respectively.}
    \label{fig:MIR_emission_WR_candidates}
\end{figure}

\begin{figure}[t]
    \centering
        \includegraphics[width=0.47\columnwidth]{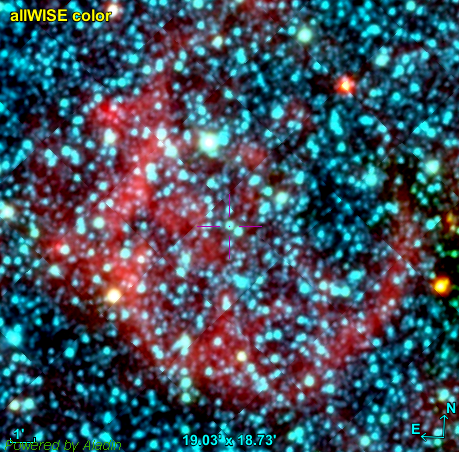}
    \hspace{0.005\textwidth} 
        \includegraphics[width=0.47\columnwidth]{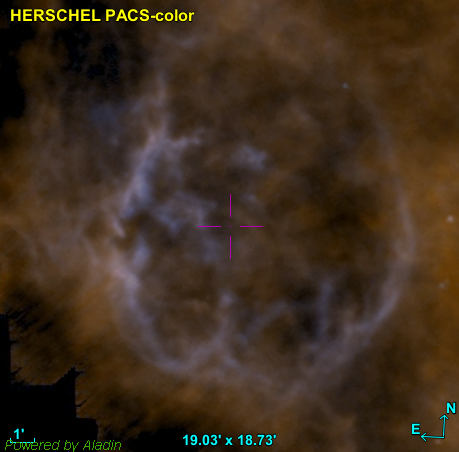}
    \hspace{0.005\textwidth}
        \includegraphics[width=0.47\columnwidth]{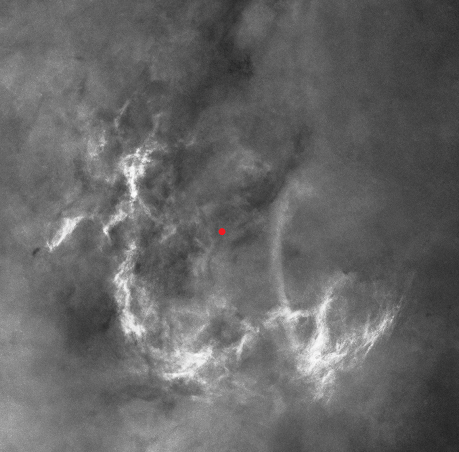}
    \hspace{0.005\textwidth}
        \includegraphics[width=0.47\columnwidth]{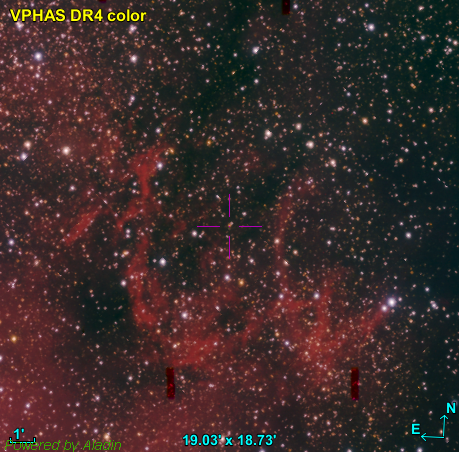}
    \caption{Possible WR ring nebula of WR-C-35. \textbf{Upper row:} WISE color image (left) and Herschel PACS color image (right). \textbf{Bottom row:} Starless H$\alpha$ image processed from SHS (left), VPHAS+ image from \textit{u, g, r, i,} H$\alpha$ bands (right).}
    \label{fig:Possible_WR_Ring_Neb}
\end{figure}

Two of our candidates, WR-C-27 and WR-C-35, show mid-IR emission in their vicinity (cf. Figs.~\ref{fig:MIR_emission_WR_candidates} and \ref{fig:Possible_WR_Ring_Neb}). WR-C-27 exhibits an arc-shaped emission in the WISE \textit{W4} band, but there is no obvious optical counterpart in the SuperCOSMOS H$\alpha$ Survey \citep[SHS;][]{2005MNRAS.362..689P} or the  VST Photometric H$\alpha$ Survey of the Southern Galactic Plane and Bulge \citep[VPHAS+;][]{2014MNRAS.440.2036D}. The case of WR-C-35 is more interesting: the mid-IR emission has a bright optical counterpart clearly visible in the SHS survey. The nebula is filamentary in shape, with an approximate apparent size of 16$\times$13 arcmin (cf. Fig.~\ref{fig:Possible_WR_Ring_Neb}). Until recently, this nebula was identified as a supernova remnant (SNR) named G11.1-1.0, but it was removed from the 2024 version of the SNRs catalog \citep{green2024updatedcatalogue310galactic} after several authors instead identified this object as an HII region \citep{2019A&A...623A.105G, 2021A&A...651A..86D}. WR-C-35 is located near the center of the nebula, which may actually be a new WR ring nebula. Further analysis that is outside the scope of this paper is required to confirm whether this nebula is indeed associated with the WR star.

 %--------------------------------------------------------------------
\section{Conclusion}
\label{sec:Conclusion}

The application of the ELP-ELS algorithm to the \textit{Gaia} DR3 data, specifically low-resolution XP spectra, resulted in the identification of 565 sources as potential WR candidates that were published as part of the data release. Approximately half of these were already known Galactic or Magellanic WR stars. According to our investigation, about 40\% of the sample are likely contaminants, with faint magnitudes and unreliable pEW H$\alpha$ measurements. Our selection criteria, primarily based on 2MASS colors and pEW H$\alpha$ measurements, produced a refined list of 59 strong candidates ($\sim$10\% of the original sample).

In this study, we focused our spectroscopic observations on 37 candidates brighter than 
$G = 16$ mag. Spectra were acquired using small telescopes in Chile and France, as well as the 1-m C2PU's Epsilon telescope at the Calern Observatory. We spectroscopically confirmed 33 new Galactic WR stars, comprising 16 WN and 17 WC-type stars. This total includes three WR stars independently discovered by \cite{2024AJ....168..167M}. Of the observed candidates, only three were found not to be WR stars.

Additionally, we report the detection of a possible new WR ring nebula surrounding WR-C-35, which we confirmed to be a WN8 star. This WR star lies near the center of a complex filamentary optical nebula, typical of material shaped by fast, hot stellar winds. Previously identified as an SNR under the designation G11.1-1.0, the nebula was recently reclassified as an HII region by \cite{2019A&A...623A.105G} and \cite{2021A&A...651A..86D}.

We also identified discrepancies in the coordinates of some objects in both the online GWRC and the SIMBAD database. The correct coordinates for WR31-1 (THA 35-II-153) are given by SIMBAD, while the GWRC lists incorrect values. WR111-6 has inaccurate coordinates in both sources; they should correspond to Gaia DR3 4095702533139975168. The GWRC incorrectly identifies HD 158860 as WR94 instead of CD-33 1268B. For the Magellanic WR star Brey 11, SIMBAD incorrectly refers to HD 268856, whereas the correct identification is Gaia DR3 4655312994920412416.

Recent discoveries, including those by \cite{2024AJ....168..167M} and our own work, demonstrate that undiscovered Galactic WR stars remain to be found. The \textit{Gaia} low-resolution XP spectra provide a valuable option for this effort. Our starting sample was limited to the ESP-ELS algorithm outputs, which only include a subset of candidate H$\alpha$ emitters. Future work will aim to expand the analysis to a broader sample, applying the insights gained from this study. In a similar way, we are currently conducting a search for new symbiotic systems (Merc et al., in prep.).

In a subsequent paper of this series, we will present follow-up results for our fainter WR candidates observed with larger-diameter telescopes. Additionally, we will analyze the photometric data available for our candidates to investigate variability and potential binarity.

\begin{acknowledgements}
We are thankful to an anonymous referee for carefully reading the manuscript and for the comments and suggestions that improved it. We warmly thank Jean-Pierre Rivet from the Calern Observatory for granting us access to C2PU's Epsilon telescope. We also thank Marcel Drechsler for processing the starless version image of the possible WR-C-35 ring nebula from SHS. The research of J.M. was supported by the Czech Science Foundation (GACR) project no. 24-10608O and by the Spanish Ministry of Science and Innovation with grant no. PID2023-146453NB-100 (PLAtoSOnG).

This work has made use of data from the European Space Agency (ESA) mission
{\it Gaia} (\url{https://www.cosmos.esa.int/gaia}), processed by the {\it Gaia}
Data Processing and Analysis Consortium (DPAC,
\url{https://www.cosmos.esa.int/web/gaia/dpac/consortium}). Funding for the DPAC
has been provided by national institutions, in particular the institutions
participating in the {\it Gaia} Multilateral Agreement. This work has also made use of the Python package GaiaXPy, developed and maintained by members of the Gaia Data Processing and Analysis Consortium (DPAC), and in particular, Coordination Unit 5 (CU5), and the Data Processing Centre located at the Institute of Astronomy, Cambridge, UK (DPCI). This publication has made use of data products from the Two Micron All Sky Survey, which is a joint project of the University of Massachusetts and the Infrared Processing and Analysis Center/California Institute of Technology, funded by the National Aeronautics and Space Administration and the National Science Foundation.
This research has used SIMBAD database \citep{2000A&AS..143....9W}, Vizier catalog access tool \citep{2000A&AS..143...23O}, Aladin Sky Atlas \citep{2000A&AS..143...33B}, and the cross-match service \citep{2012ASPC..461..291B}  provided by CDS, Strasbourg Observatory, France. This research has made use of the SVO Filter Profile Service "Carlos Rodrigo", funded by MCIN/AEI/10.13039/501100011033/ through grant PID2020-112949GB-I00.

\end{acknowledgements}

   \bibliographystyle{aa} % style aa.bst
  \bibliography{bibliography.bib} % your references Yourfile.bib

\begin{appendix}

\begin{figure*}[!h]
\section{Optical spectra of our candidates}
   \centering
   \includegraphics [width=0.9\textwidth] {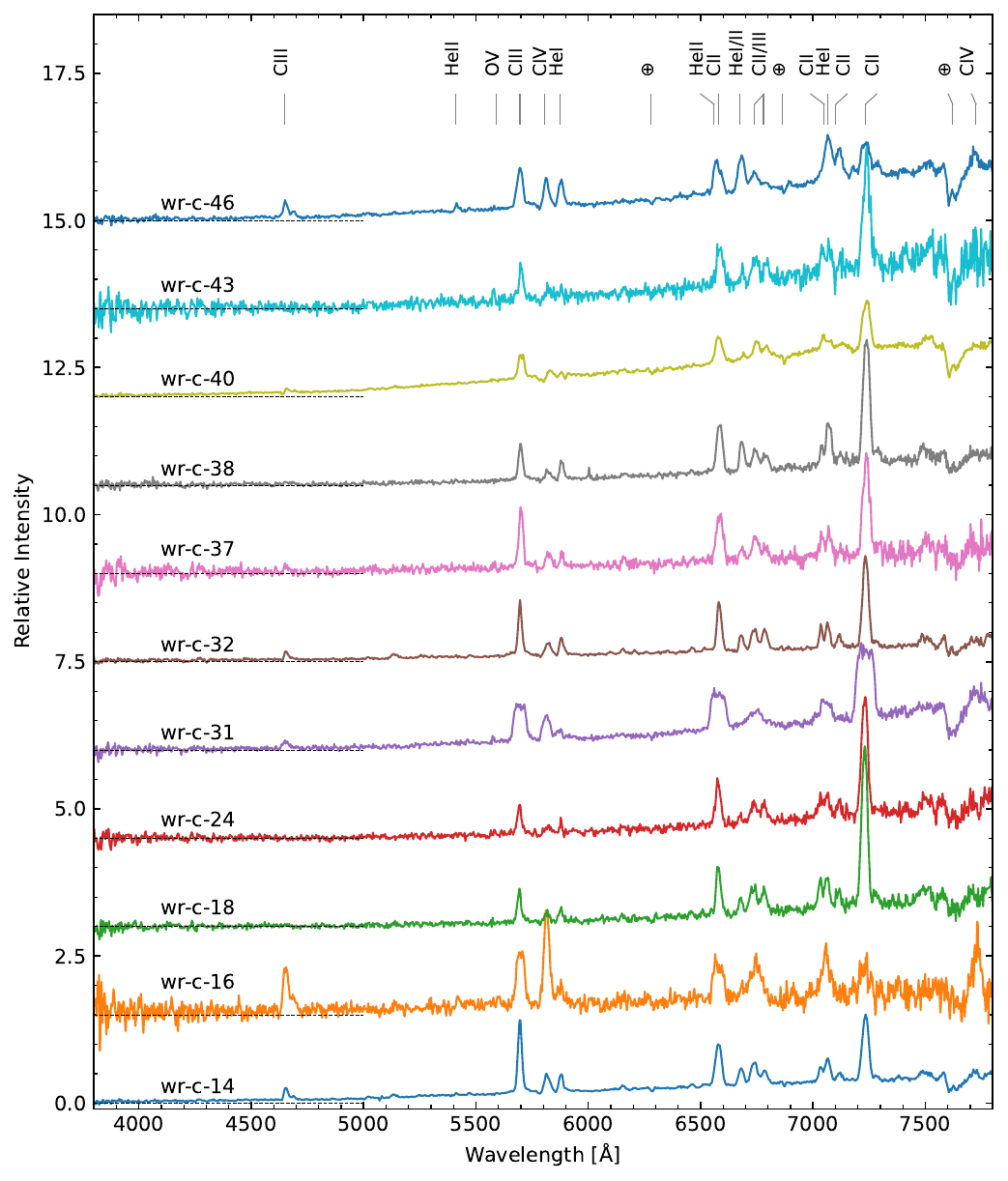}
   \caption{Spectra of newly discovered WC8 and WC9 stars. Spectra are dominated by \ion{C}{iii} emission. WR-C-31, WR-C-16 and WR-C-46 are the only WC8 stars in the sample, while the other stars belong to the WC9 subclass. The black dashed lines indicate the zero intensity scale of each spectrum. The position of the telluric absorption bands is marked with the symbol $\oplus$.}
   \label{fig:WCL}%
\end{figure*}

\begin{figure*}[!h]
   \centering
   \includegraphics [width=0.9\textwidth] {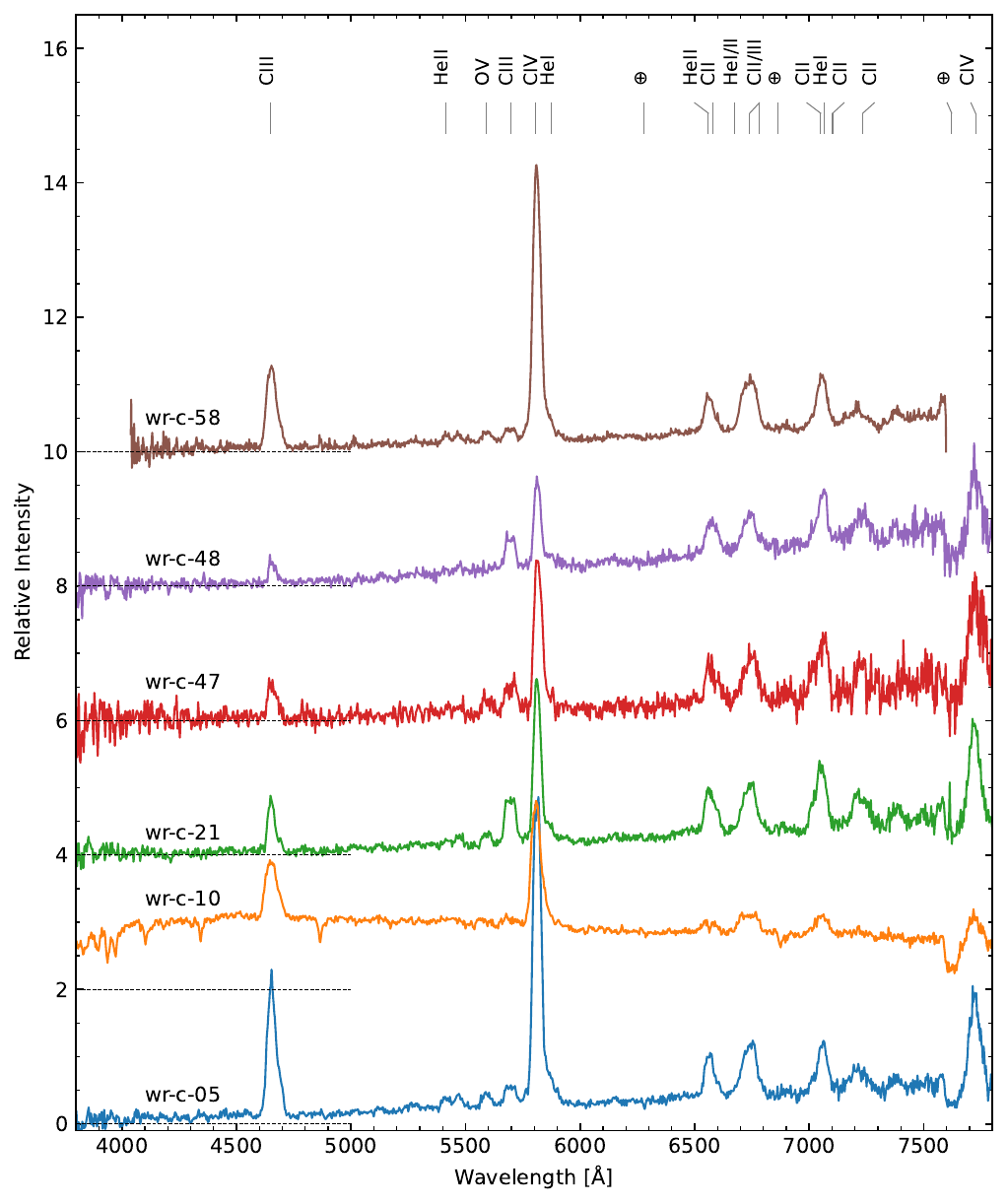}
   \caption{Spectra of newly discovered WC5, WC6 and WC7 stars. Spectra are dominated by \ion{C}{iv} emission. According to the relative intensity of their \ion{O}{v} and \ion{C}{iii} lines, WR-C-05, WR-C-10, and WR-C-58 belong to the WC5 subclass, WR-C-47 to WC6, while all other stars are WC7. The black dashed lines indicate the zero intensity scale of each spectrum. The position of the telluric absorption bands is marked with the symbol $\oplus$.}
   \label{fig:WC5_7}%
\end{figure*}

%--------------------------------------------------------------------
\begin{figure*}[!h]
   \centering
   \includegraphics [width=0.9\textwidth] {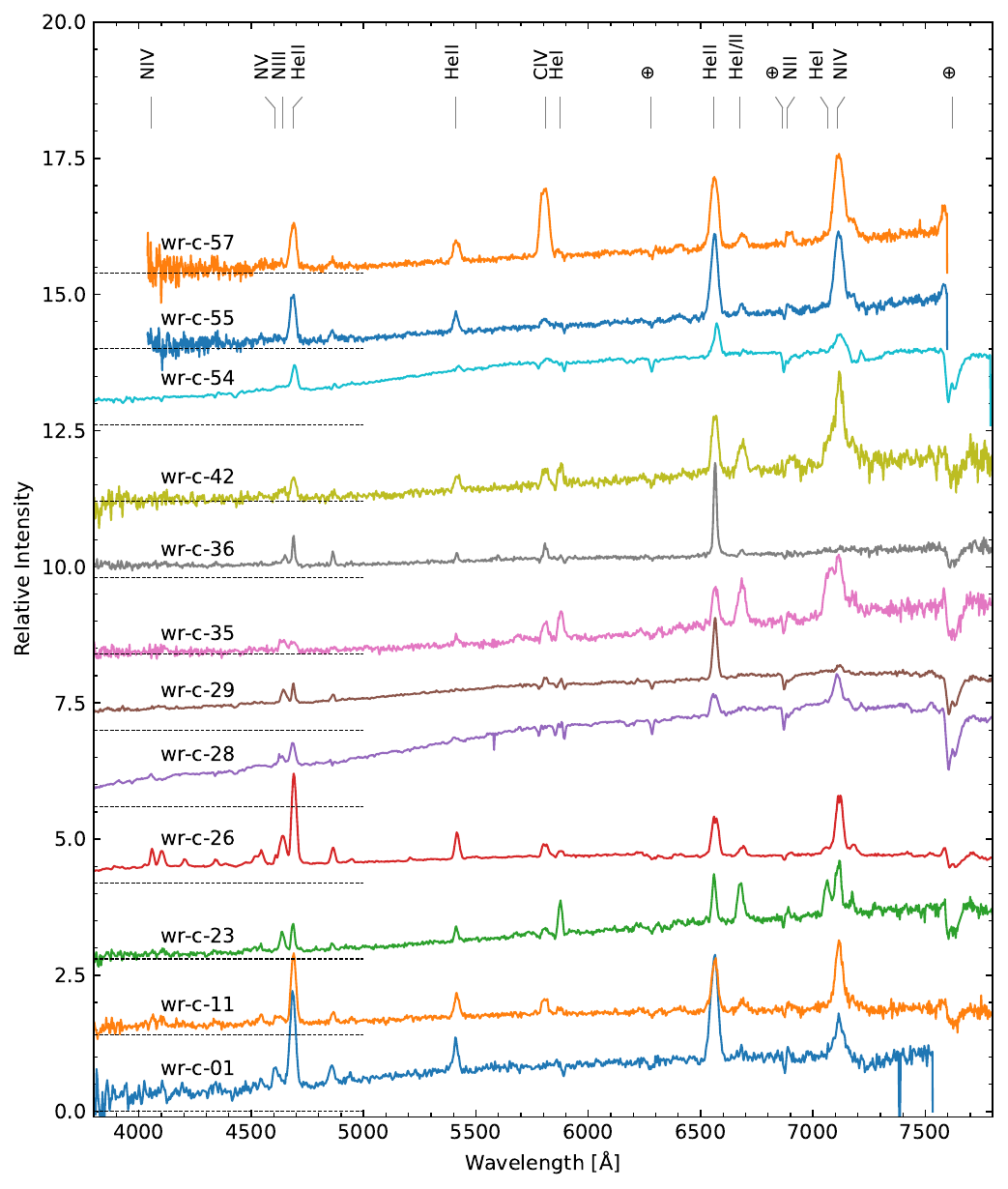}
   \caption{Spectra of newly discovered WN stars. WN/C: WR-C-57 for its unusually strong \ion{C}{iv} $\lambda$5804 line, WN3: WR-C-55, WN4: WR-C-01, WN5: WR-C-11, WN6: WR-C-26 (already known as WR94, see text), WR-C-28, WR-C-42, WR-C-54, WN7: WR-C-29, WR-C-36, WN8: WR-C-23. The black dashed lines indicate the zero intensity scale of each spectrum. The position of the telluric absorption bands is marked with the symbol $\oplus$.}
   \label{fig:WN_classified}%
\end{figure*}

\begin{figure*}[!h]
   \centering
   \includegraphics [width=0.9\textwidth] {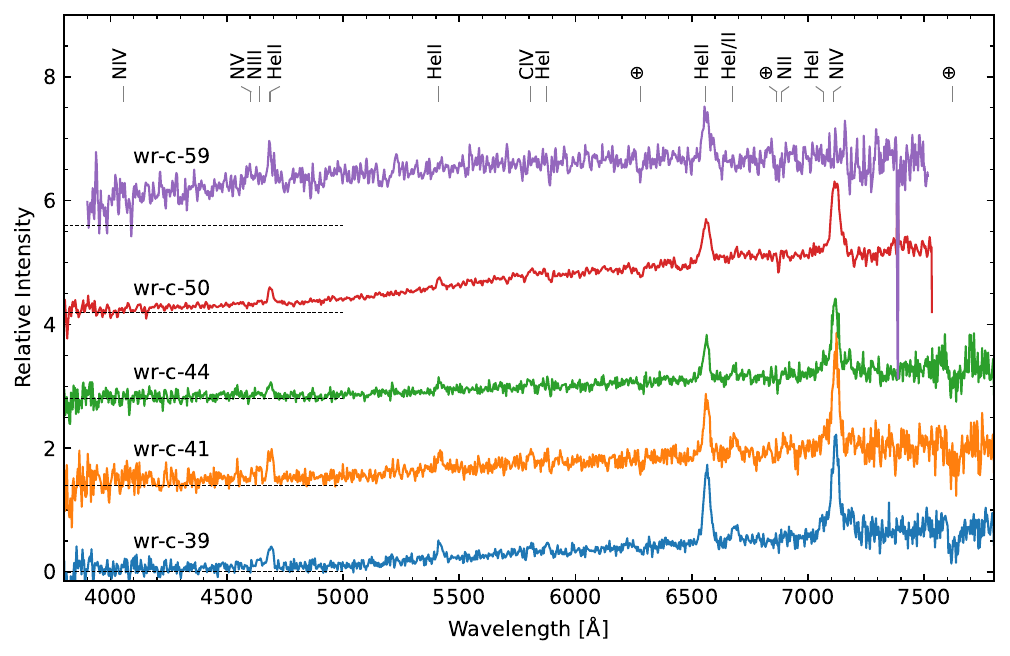}
   \caption{Possible WN stars for which we have only low S/N spectra available. WR-C-59 likely belongs to the WN3 subclass, WR-C-44, WR-C-41 and WR-C-50 to the range of subclasses WN5-6, and WR-C-39 to the range of subclasses WN7-8. The black dashed lines indicate the zero intensity scale of each spectrum. The position of the telluric absorption bands is marked with the symbol $\oplus$.}
   \label{fig:WN_likely}%
\end{figure*}

\begin{figure*}[!h]
   \centering
   \includegraphics [width=0.9\textwidth] {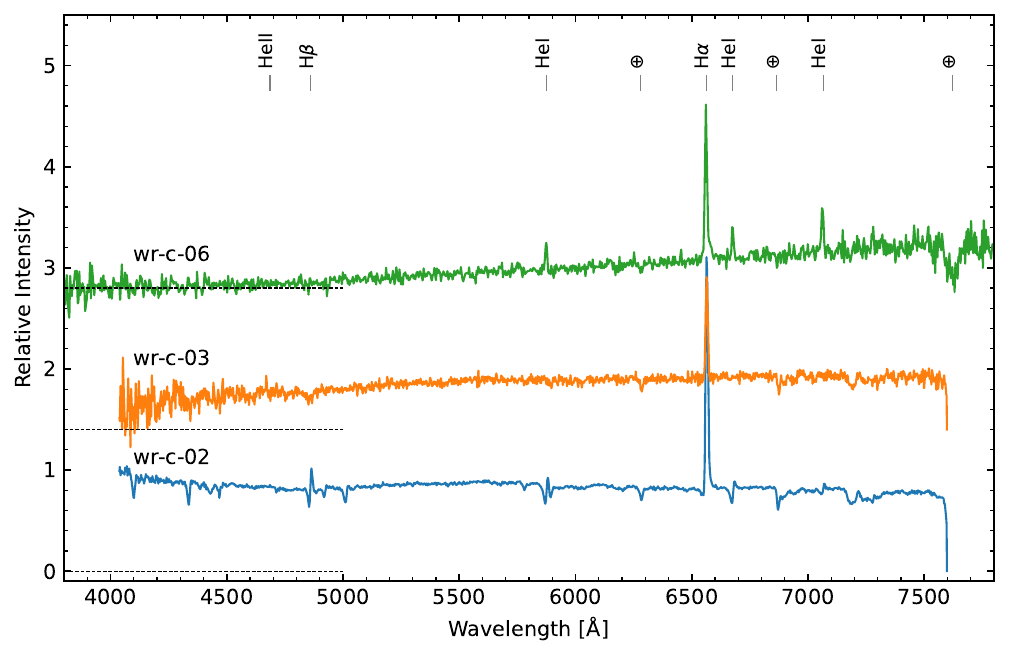}
   \caption{Observed candidates with non-WR spectra.The black dashed lines indicate the zero intensity scale of each spectrum. The position of the telluric absorption bands is marked with the symbol $\oplus$.}
   \label{fig:uncertain_nature}%
\end{figure*}

\newpage\onecolumn
\section{Additional tables}
\begin{table}[h!]
    \centering
    \renewcommand{\arraystretch}{1.01}
    \caption{Log of spectroscopic observations of WR stars candidates, together with their classification.}
    \begin{tabular}[t]{cccccc}
    \hline\hline
Object & Exposure time & Julian Date & Obs. Site & Subclass \\ \hline
WR-C-01 & 8 x 800 s & 2460621.3142 & Kermerrien & WN4 \\
WR-C-02 & 3 x 180 s & 2460613.6058 & C2PU & Not WR \\
WR-C-03 & 2 x 1200 s & 2460614.5568 & C2PU & Not WR \\
WR-C-05 & 4 x 1200 s & 2460618.8110 & DSC & WC5 \\
WR-C-06 & 6 x 1200 s & 2460628.8177 & DSC & Not WR \\
WR-C-10 & 6 x 1200 s & 2460642.7981 & DSC & WC5+abs \\
WR-C-11 & 4 x 1200 s & 2460651.8212 & DSC & WN5 \\
WR-C-14 & 5 x 1200 s & 2460535.5095 & DSC & WC9 \\
WR-C-16 & 6 x 1200 s & 2460555.5337 & DSC & WC8 \\
WR-C-18 & 4 x 1200 s & 2460549.5357 & DSC & WC9d \\
WR-C-21 & 5 x 1200 s & 2460544.5909 & DSC & WC7 \\
WR-C-23 & 4 x 1200 s & 2460544.5182 & DSC & WN8 \\
WR-C-24 & 4 x 1200 s & 2460554.5267 & DSC & WC9d \\
WR-C-26 & 2 x 1200 s & 2460548.5107 & DSC & WN6 (WR94) \\
WR-C-28 & 3 x 1200 s & 2460548.5512 & DSC & WN6+abs \\
WR-C-29 & 4 x 1200 s & 2460546.6710 & DSC & WN7 \\
WR-C-31 & 4 x 1200 s & 2460547.6661 & DSC & WC8 \\
WR-C-32 & 3 x 1200 s & 2460544.6543 & DSC & WC9d \\
WR-C-35 & 2 x 1200 s & 2460547.7168 & DSC & WN8 \\
WR-C-36 & 4 x 1200 s & 2460549.5999 & DSC & WN7h \\
WR-C-37 & 6 x 1200 s & 2460559.5633 & DSC & WC9 \\
WR-C-38 & 4 x 1200 s & 2460548.6088 & DSC & WC9 \\
WR-C-39 & 4 x 1200 s & 2460546.6092 & DSC & WN7-8 \\
WR-C-40 & 3 x 1200 s & 2460554.5815 & DSC & WC9 \\
WR-C-41 & 5 x 1200 s & 2460559.6487 & DSC & WN5-6 \\
WR-C-42 & 7 x 1200 s & 2460555.6315 & DSC & WN6 \\
WR-C-43 & 5 x 1200 s & 2460548.6809 & DSC & WC9d \\
WR-C-44 & 5 x 1200 s & 2460554.6455 & DSC & WN5-6 \\
WR-C-46 & 5 x 1200 s & 2460549.6716 & DSC & WC8 \\
WR-C-47 & 6 x 1200 s & 2460561.0857 & DSC & WC6 \\
WR-C-48 & 4 x 1200 s & 2460561.5978 & DSC & WC7 \\
WR-C-50 & 10 x 1200 s & 2460549.4637 & Kermerrien & WN5-6 \\
WR-C-54 & 3 x 1200 s & 2460626.2761 & Cornillon & WN6+abs \\
WR-C-55 & 4 x 1200 s & 2460614.4183 & C2PU & WN3 \\
WR-C-57 & 4 x 1200 s & 2460615.3789 & C2PU & WN/C \\
WR-C-58 & 3 x 1200 s & 2460613.3071 & C2PU & WC5 \\
WR-C-59 & 7 x 1200 s & 2460626.4319 & Kermerrien & WN3 \\
\hline
    \end{tabular}
    \label{tab:Observed_WR_Candidates}
\end{table}

\newpage
{\setlength{\LTcapwidth}{\dimexpr1\textwidth}
\centering

\renewcommand{\arraystretch}{1.4}
\begin{longtable}{cccccccc}
\caption{\textit{Gaia} DR3 \citep{2023A&A...674A...1G} parallax $\varpi$, goodness-of-fit (GOF), RUWE of target stars. The geometric distances from \citet{2021AJ....161..147B} are listed and adopted for the calculations in this work. The calculated vertical distances
 from the Galactic disc are also included in the Table. The last column contains the subclass of each target. \label{tab:paral}}\\

\hline\hline
Name & $\varpi$ [mas] & $\varpi$/$\sigma_\varpi$ & RUWE$^a$ & GOF$^b$ & Dist. (B-J) [kpc] & z [kpc] & Subclass\\
\hline
\endfirsthead
\caption{continued.}\\
\hline\hline
Name & $\varpi$ [mas] & $\varpi$/$\sigma_\varpi$ & RUWE & GOF & Dist. (B-J) [kpc] & z [kpc] & Subclass\\
\hline
\endhead
\hline
\endfoot
WR-C-01 & 0.131$\pm$0.014 & 9.4 & 1.06 & 1.96 & 6.11$^{+0.57}_{-0.47}$ & 144$^{+13}_{-11}$ & WN4 \\
WR-C-02 & 0.261$\pm$0.016 & 16.0 & 1.03 & 0.62 & 3.46$^{+0.25}_{-0.18}$ & 32$^{+2}_{-2}$ & Not WR \\
WR-C-03 & 0.105$\pm$0.042 & 2.5 & 1.08 & 1.74 & 6.50$^{+1.75}_{-1.27}$ & -348$^{+68}_{-93}$ & Not WR \\
WR-C-04 & 0.137$\pm$0.041 & 3.4 & 0.98 & -0.54 & 6.14$^{+2.17}_{-1.07}$ & -78$^{+14}_{-28}$ & .. \\
WR-C-05 & -0.155$\pm$0.067 & - & 3.86 & 53.56 & 11.85$^{+3.56}_{-3.09}$ & -270$^{+70}_{-81}$ & WC5 \\
WR-C-06 & 0.147$\pm$0.024 & 6.0 & 1.05 & 1.26 & 4.97$^{+0.72}_{-0.50}$ & -37$^{+4}_{-5}$ & Not WR \\
WR-C-07 & 0.029$\pm$0.045 & 0.6 & 1.20 & 5.69 & 8.14$^{+2.23}_{-1.47}$ & -120$^{+22}_{-33}$ & .. \\
WR-C-08 & 0.161$\pm$0.082 & 2.0 & 1.21 & 5.25 & 4.53$^{+1.62}_{-1.02}$ & -106$^{+24}_{-38}$ & .. \\
WR-C-09 & 0.114$\pm$0.040 & 2.9 & 1.01 & 0.33 & 6.26$^{+1.37}_{-1.15}$ & 5$^{+1}_{-1}$ & .. \\
WR-C-10 & 0.156$\pm$0.022 & 7.3 & 1.20 & 5.21 & 5.09$^{+0.57}_{-0.58}$ & -34$^{+4}_{-4}$ & WC5+abs \\
WR-C-11 & 0.056$\pm$0.019 & 2.9 & 0.98 & -0.56 & 10.31$^{+2.10}_{-1.55}$ & -338$^{+51}_{-69}$ & WN5 \\
WR-C-12 & 0.196$\pm$0.049 & 4.0 & 0.99 & -0.25 & 4.56$^{+1.13}_{-0.80}$ & -33$^{+6}_{-8}$ & .. \\
WR-C-13 & 0.224$\pm$0.074 & 3.0 & 1.01 & 0.36 & 4.01$^{+1.27}_{-0.88}$ & -43$^{+9}_{-13}$ & .. \\
WR-C-14 & 0.258$\pm$0.023 & 11.2 & 1.76 & 22.87 & 3.26$^{+0.27}_{-0.23}$ & 26$^{+2}_{-2}$ & WC9 \\
WR-C-15 & -0.022$\pm$0.083 & - & 1.00 & 0.11 & 6.13$^{+1.91}_{-1.27}$ & -89$^{+18}_{-28}$ & .. \\
WR-C-16 & 0.099$\pm$0.043 & 2.3 & 1.17 & 5.13 & 6.50$^{+2.08}_{-1.48}$ & -201$^{+46}_{-64}$ & WC8 \\
WR-C-17 & -0.062$\pm$0.056 & - & 1.05 & 1.32 & 9.93$^{+2.57}_{-2.11}$ & 881$^{+228}_{-187}$ & .. \\
WR-C-18 & 0.193$\pm$0.054 & 3.6 & 1.12 & 3.22 & 3.74$^{+0.96}_{-0.62}$ & -30$^{+5}_{-8}$ & WC9d \\
WR-C-19 & 0.037$\pm$0.050 & 0.7 & 1.01 & 0.40 & 8.35$^{+2.77}_{-1.80}$ & -211$^{+45}_{-70}$ & .. \\
WR-C-20 & 1.954$\pm$0.188 & 10.4 & 2.84 & 45.67 & 0.51$^{+0.03}_{-0.04}$ & -6$^{+0}_{-0}$ & .. \\
WR-C-21 & 0.144$\pm$0.029 & 5.0 & 1.07 & 1.86 & 4.98$^{+0.66}_{-0.56}$ & 125$^{+17}_{-14}$ & WC7 \\
WR-C-22 & 0.075$\pm$0.109 & 0.7 & 1.23 & 6.17 & 5.64$^{+2.36}_{-1.84}$ & -131$^{+43}_{-55}$ & .. \\
WR-C-23 & 0.203$\pm$0.018 & 11.5 & 1.03 & 0.79 & 3.95$^{+0.24}_{-0.24}$ & -197$^{+12}_{-12}$ & WN8 \\
WR-C-24 & 0.180$\pm$0.053 & 3.4 & 1.09 & 2.74 & 4.06$^{+0.82}_{-0.72}$ & -2$^{+0}_{-0}$ & WC9d \\
WR-C-25 & 0.165$\pm$0.129 & 1.3 & 1.09 & 2.11 & 5.76$^{+3.83}_{-1.95}$ & -99$^{+33}_{-66}$ & .. \\
WR-C-26 & 0.413$\pm$0.021 & 19.4 & 0.92 & -1.57 & 2.30$^{+0.11}_{-0.11}$ & -10$^{+0}_{-0}$ & WN6 \\
WR-C-27 & 0.090$\pm$0.049 & 1.8 & 0.98 & -0.43  & 6.05$^{+1.93}_{-1.24}$ & -36$^{+7}_{-12}$ & .. \\
WR-C-28 & 0.272$\pm$0.018 & 14.9 & 1.08 & 1.92 & 3.22$^{+0.20}_{-0.20}$ & -20$^{+1}_{-1}$ & WN6+abs \\
WR-C-29 & 0.395$\pm$0.018 & 21.5 & 0.87 & -2.87 & 2.39$^{+0.10}_{-0.11}$ & -35$^{+2}_{-1}$ & WN7 \\
WR-C-30 & -0.135$\pm$0.133 & - & 1.09 & 2.90 & 7.83$^{+4.31}_{-2.68}$ & 9$^{+5}_{-3}$ & .. \\
WR-C-31 & 0.124$\pm$0.028 & 4.4 & 1.06 & 1.46 & 5.33$^{+1.00}_{-0.61}$ & 110$^{+21}_{-13}$ & WC8 \\
WR-C-32 & 0.240$\pm$0.030 & 7.9 & 0.89 & -2.33 & 3.99$^{+0.58}_{-0.45}$ & 30$^{+4}_{-3}$ & WC9d \\
WR-C-33 & 0.120$\pm$0.077 & 1.6 & 0.96 & -0.78 & 6.20$^{+2.55}_{-1.59}$ & -215$^{+55}_{-88}$ & .. \\
WR-C-34 & 0.253$\pm$0.072 & 3.5 & 1.09 & 1.75 & 3.70$^{+1.25}_{-0.77}$ & 4$^{+1}_{-1}$ & .. \\
WR-C-35 & 0.171$\pm$0.027 & 6.4 & 1.12 & 2.46 & 4.24$^{+0.51}_{-0.36}$ & -80$^{+7}_{-10}$ & WN8 \\
WR-C-36 & 0.297$\pm$0.025 & 11.8 & 1.05 & 1.08 & 2.94$^{+0.17}_{-0.19}$ & 48$^{+3}_{-3}$ & WN7h \\
WR-C-37 & 0.116$\pm$0.045 & 2.6 & 1.08 & 1.63 & 5.71$^{+1.37}_{-1.12}$ & 402$^{+97}_{-79}$ & WC9 \\
WR-C-38 & 0.179$\pm$0.043 & 4.2 & 0.91 & -1.91 & 4.22$^{+0.93}_{-0.70}$ & 39$^{+9}_{-6}$ & WC9 \\
WR-C-39 & 0.128$\pm$0.028 & 4.6 & 1.02 & 0.50 & 5.44$^{+0.90}_{-0.60}$ & 110$^{+18}_{-12}$ & WN7-8 \\
WR-C-40 & 0.211$\pm$0.064 & 3.3 & 3.18 & 36.63 & 4.33$^{+1.33}_{-0.73}$ & 13$^{+4}_{-2}$ & WC9 \\
WR-C-41 & 0.045$\pm$0.040 & 1.1 & 0.99 & -0.26 & 8.22$^{+2.38}_{-1.62}$ & -290$^{+57}_{-84}$ & WN5-6 \\
WR-C-42 & 0.035$\pm$0.035 & 1.0 & 1.06 & 1.19 & 8.35$^{+1.93}_{-1.52}$ & 191$^{+44}_{-35}$ & WN6 \\
WR-C-43 & 0.199$\pm$0.053 & 3.8 & 1.15 & 3.14 & 3.63$^{+0.56}_{-0.57}$ & 6$^{+1}_{-1}$ & WC9d \\
WR-C-44 & 0.165$\pm$0.037 & 4.5 & 1.09 & 1.84 & 5.11$^{+1.00}_{-0.86}$ & -129$^{+22}_{-25}$ & WN5-6 \\
WR-C-45 & 0.182$\pm$0.108 & 1.7 & 1.12 & 2.46 & 4.27$^{+1.50}_{-1.21}$ & -100$^{+28}_{-35}$ & .. \\
WR-C-46 & 0.242$\pm$0.031 & 7.8 & 1.32 & 5.77 & 3.22$^{+0.35}_{-0.27}$ & 67$^{+7}_{-6}$ & WC8 \\
WR-C-47 & 0.140$\pm$0.048 & 2.9 & 1.10 & 2.36 & 5.46$^{+1.56}_{-1.10}$ & 152$^{+43}_{-30}$ & WC6 \\
WR-C-48 & -1.902$\pm$0.452 & - & 13.84 & 251.32 & 5.93$^{+1.71}_{-1.67}$ & 21$^{+6}_{-6}$ & WC7 \\
WR-C-49 & - & - & - & 669.89 & - & - & .. \\
WR-C-50 & 0.206$\pm$0.142 & 1.4 & 9.60 & 162.57  & 4.49$^{+2.15}_{-1.36}$ & -20$^{+6}_{-9}$ & WN5-6 \\
WR-C-51 & -0.077$\pm$0.083 & - & 0.92 & -3.11 & 9.99$^{+3.83}_{-2.79}$ & 199$^{+76}_{-55}$ & .. \\
WR-C-52 & 0.101$\pm$0.047 & 2.2 & 1.03 & 0.87 & 6.72$^{+1.71}_{-1.25}$ & -227$^{+42}_{-58}$ & .. \\
WR-C-53 & 0.122$\pm$0.045 & 2.7 & 0.98 & -0.53 & 6.44$^{+2.14}_{-1.29}$ & 222$^{+74}_{-44}$ & .. \\
WR-C-54 & 0.271$\pm$0.015 & 17.7 & 1.43 & 11.72 & 3.36$^{+0.17}_{-0.15}$ & 86$^{+4}_{-4}$ & WN6+abs \\
WR-C-55 & 0.099$\pm$0.023 & 4.2 & 1.04 & 1.05 & 6.98$^{+1.09}_{-0.86}$ & 110$^{+17}_{-14}$ & WN3 \\
WR-C-56 & -0.051$\pm$0.078 & - & 1.07 & 1.78 & 7.17$^{+2.28}_{-1.92}$ & 374$^{+119}_{-100}$ & .. \\
WR-C-57 & 0.114$\pm$0.028 & 4.0 & 1.07 & 2.06 & 6.14$^{+1.20}_{-0.86}$ & 281$^{+55}_{-40}$ & WN/C \\
WR-C-58 & 0.126$\pm$0.025 & 5.0 & 1.15 & 3.80 & 5.70$^{+0.90}_{-0.71}$ & 122$^{+19}_{-15}$ & WC5 \\
WR-C-59 & 0.084$\pm$0.023 & 3.7 & 1.00 & -0.01 & 8.69$^{+1.68}_{-1.35}$ & 533$^{+103}_{-83}$ & WN3 \\
\end{longtable}
\tablefoot{$^a$Renormalised Unit Weight Error (RUWE). A value close to 1.0 suggests a good fit with a single-star model, whereas notably higher values may point to problems with the astrometric solution or the presence of a binary companion. $^b$Goodness-of-fit statistic for the astrometric solution. As per the \textit{Gaia} DR3 documentation, values of $\gtrsim$3 indicate a poor fit to the observed data.
}}

\vspace{6mm}

{\setlength{\LTcapwidth}{\dimexpr1\textwidth}
\centering
%\small 
%\setlength{\tabcolsep}{5.8pt}
\renewcommand{\arraystretch}{1.4}
\begin{longtable}{ccccccccc}
\caption{Absolute M$_v$ magnitudes of selected WR candidates. The distances from  \citet{2021AJ....161..147B} were used in all the calculations (see Table \ref{tab:paral}).\label{tab:absolute_mags}}\\

\hline\hline
Name & \textit{v} & $(b-v)$$^{a}$ & $(b-v)$$^{b}$ & ${A_v}^{a}$ & ${A_v}^{b}$ & ${M_v}^{a}$ & ${M_v}^{b}$ & Subclass \\
 & {[}mag{]} & {[}mag{]} & {[}mag{]} & {[}mag{]} & {[}mag{]} & {[}mag{]} & {[}mag{]} & \\
\hline
\endfirsthead
\caption{continued.}\\
\hline\hline
Name & \textit{v} & $(b-v)$$^{a}$ & $(b-v)$$^{b}$ & ${A_v}^{a}$ & ${A_v}^{b}$ & ${M_v}^{a}$ & ${M_v}^{b}$ & Subclass \\
 & {[}mag{]} & {[}mag{]} & {[}mag{]} & {[}mag{]} & {[}mag{]} & {[}mag{]} & {[}mag{]} & \\
\hline
\endhead
\hline
\endfoot
WR-C-01 & 14.96$\pm0.01$ & 1.08$\pm0.03$ & 1.30$\pm0.11$ & 5.8$\pm0.4$ & 6.7$\pm0.6$ & -4.7$\pm0.5$ & -5.7$\pm0.6$ & WN4\\
WR-C-02 & 11.43$\pm0.02$ & 0.57$\pm0.03$ & 0.38$\pm0.02$ & $^{\rm av}$ 3.6$\pm0.4$ & $^{\rm av}$ 2.8$\pm0.4$ & -4.9$\pm0.5$ & -4.1$\pm0.4$ & Not WR\\
WR-C-03 & 16.34$\pm0.02$ & 0.63$\pm0.05$ & - & $^{\rm av}$ 3.8$\pm0.5$ & - & -1.6$\pm0.7$ & - & Not WR\\
%WR-C-04 & - & - & - & - & - & - & - \\
WR-C-05 & 16.57$\pm0.07$ & 1.60$\pm0.17$ & - & 7.4$\pm1.1$ & - & -6.2$\pm1.2$ & - & WC5\\
WR-C-06 & 17.22$\pm0.04$ & 1.46$\pm0.15$ & - & $^{\rm av}$ 7.3$\pm0.7$ & - & -3.5$\pm0.8$ & - & Not WR\\
%WR-C-07 & - & - & - & - & - & - & - \\
%WR-C-08 & - & - & - & - & - & - & - \\
%WR-C-09 & - & - & - & - & - & - & - \\
WR-C-10 & 16.88$\pm0.11$ & 1.27$\pm0.35$ & - & 6.1$\pm1.7$ & - & -2.7$\pm1.7$ & - & WC5+abs \\
WR-C-11 & 15.86$\pm0.02$ & 1.03$\pm0.05$ & - & 5.4$\pm0.5$ & - & -4.6$\pm0.6$ & - & WN5 \\
%WR-C-12 & - & - & - & - & - & - & - \\
%WR-C-13 & - & - & - & - & - & - & - \\
WR-C-14 & 15.59$\pm0.02$ & 1.22$\pm0.06$ & 0.90$\pm0.14$ & 6.4$\pm0.5$ & 5.0$\pm0.7$ & -3.3$\pm0.5$ & -2.0$\pm0.7$ & WC9\\
WR-C-15 & 20.49$\pm0.27$ & 2.33:: & - & $^{\rm av}$ 10.9:: & - & -4.3::\textdagger & - & .. \\
WR-C-16 & 18.96$\pm0.10$ & 1.54$\pm0.47$ & - & 7.9$\pm2.0$ & - & -3.0$\pm2.1$ & - & WC8\\
%WR-C-17 & - & - & - & - & - & - & - \\
WR-C-18 & 19.34$\pm0.18$ & 4.02:: & - & 17.9:: & - & -11.4::\textdagger & - & WC9d\\
WR-C-19 & 19.46$\pm0.12$ & 2.99:: & - & $^{\rm av}$ 13.6:: & - & -8.7::\textdagger & - & ..\\
WR-C-20 & 21.01$\pm0.67$ & 0.96:: & - & $^{\rm av}$ 5.2:: & - & 7.3::\textdagger & - & ..\\
WR-C-21 & 17.63$\pm0.09$ & 1.48$\pm0.29$ & - & 6.9$\pm1.4$ & - & -2.8$\pm1.5$ & - & WC7\\
%WR-C-22 & - & - & - & - & - & - & - \\
WR-C-23 & 15.47$\pm0.02$ & 1.56$\pm0.07$ & 1.32$\pm0.08$ & 7.0$\pm0.5$ & 6.1$\pm0.5$ & -4.5$\pm0.5$ & -3.6$\pm0.5$ & WN8\\
WR-C-24 & 19.47$\pm0.15$ & 2.40:: & - & 11.2:: & - & -4.8::\textdagger & - & WC9d\\
WR-C-25 & 21.58$\pm0.96$ & -0.24:: & - & $^{\rm av}$ 0.2:: & - & 7.5::\textdagger & - & ..\\
WR-C-26 & 12.24$\pm0.01$ & 0.72$\pm0.03$ & 0.73$\pm0.01$ & 4.1$\pm0.4$ & 4.1$\pm0.4$ & -3.7$\pm0.4$ & -3.7$\pm0.4$ & WN6\\
WR-C-27 & 19.74$\pm0.22$ & 2.93:: & - & $^{\rm av}$ 13.3:: & - & -7.5::\textdagger & - & ..\\
WR-C-28 & 11.99$\pm0.01$ & 1.02$\pm0.01$ & 0.99$\pm0.02$ & 5.4$\pm0.4$ & 5.2$\pm0.4$ & -5.9$\pm0.4$ & -5.8$\pm0.4$ & WN6+abs\\
WR-C-29 & 12.60$\pm0.01$ & 0.88$\pm0.02$ & 0.89$\pm0.02$ & 4.3$\pm0.4$ & 4.3$\pm0.4$ & -3.5$\pm0.4$ & -3.6$\pm0.4$ & WN7\\
%WR-C-30 & - & - & - & - & - & - & - \\
WR-C-31 & 18.81$\pm0.19$ & 1.44$\pm0.56$ & - & 7.5$\pm2.4$ & - & -2.3$\pm2.4$ & - & WC8\\
WR-C-32 & 14.86$\pm0.03$ & 1.42$\pm0.07$ & 0.92$\pm0.25$ & 7.2$\pm0.5$ & 5.1$\pm1.1$ & -5.3$\pm0.6$ & -3.3$\pm1.2$ & WC9d\\
%WR-C-33 & - & - & - & - & - & - & - \\
%WR-C-34 & - & - & - & - & - & - & - \\
WR-C-35 & 16.69$\pm0.05$ & 1.81$\pm0.21$ & - & 8.1$\pm0.9$ & - & -4.5$\pm1.0$ & - & WN8\\
WR-C-36 & 15.65$\pm0.03$ & 1.13$\pm0.08$ & - & 5.3$\pm0.5$ & - & -2.0$\pm0.5$ & - & WN7h\\
%WR-C-37 & - & - & - & - & - & - & - \\
WR-C-38 & 19.07$\pm0.19$ & 1.43$\pm0.70$ & - & 7.2$\pm2.9$ & - & -1.3$\pm2.9$ & - & WC9\\
WR-C-39 & 17.30$\pm0.06$ & 1.53$\pm0.28$ & - & 6.9$\pm1.2$ & - & -3.3$\pm1.3$ & - & WN7-8\\
WR-C-40 & 15.27$\pm0.06$ & 1.47$\pm0.40$ & 1.59$\pm0.13$ & 7.4$\pm1.7$ & 7.9$\pm0.7$ & -5.3$\pm1.8$ & -5.8$\pm0.9$ & WC9\\
%WR-C-41 & - & - & - & - & - & - & - \\
%WR-C-42 & - & - & - & - & - & - & - \\
WR-C-43 & 19.81$\pm0.30$ & 2.12:: & - & 10.1:: & - & -3.0::\textdagger & - & WC9d\\
WR-C-44 & 17.83$\pm0.07$ & 2.40$\pm0.68$ & - & 11.0$\pm2.8$ & - & -6.8$\pm2.9$ & - & WN5-6\\
%WR-C-45 & - & - & - & - & - & - & - \\
%WR-C-46 & - & - & - & - & - & - & - \\
%WR-C-47 & - & - & - & - & - & - & - \\
WR-C-48 & 18.50$\pm0.09$ & 1.44$\pm0.36$ & - & 6.8$\pm1.7$ & - & -2.1$\pm1.8$ & - & WC7\\
WR-C-49 & 17.29$\pm0.03$ & 0.48$\pm0.08$ & - & $^{\rm av}$ 3.2$\pm0.5$ & - & - & - & ..\\
WR-C-50 & 15.77$\pm0.04$ & 1.61$\pm0.10$ & 1.60$\pm0.25$ & 7.8$\pm0.6$ & 7.8$\pm1.1$ & -5.3$\pm1.0$ & -5.3$\pm1.4$ & WN5-6\\
%WR-C-51 & - & - & - & - & - & - & - \\
WR-C-52 & 18.68$\pm0.06$ & 2.05$\pm0.57$ & - & $^{\rm av}$ 9.7$\pm2.4$ & - & -5.2$\pm2.4$ & - & ..\\
%WR-C-53 & - & - & - & - & - & - & - \\
WR-C-54 & 11.20$\pm0.01$ & 0.90$\pm0.02$ & 0.91$\pm0.02$ & 4.8$\pm0.4$ & 4.9$\pm0.4$ & -6.3$\pm0.4$ & -6.3$\pm0.4$ & WN6+abs\\
WR-C-55 & 17.40$\pm0.04$ & 1.42$\pm0.12$ & - & 7.2$\pm0.6$ & - & -4.0$\pm0.7$ & - & WN3\\
%WR-C-56 & - & - & - & - & - & - & - \\
WR-C-57 & 18.65$\pm0.09$ & 1.79$\pm0.54$ & - & 8.3$\pm2.3$ & - & -3.6$\pm2.3$ & - & WN/C\\
WR-C-58 & 18.35$\pm0.12$ & 2.88:: & - & 12.7:: & - & -8.1::\textdagger & - & WC5\\
WR-C-59 & 16.02$\pm0.02$ & 0.82$\pm0.04$ & - & 4.7$\pm0.4$ & - & -3.4$\pm0.6$ & - & WN3\\
\end{longtable}}
\tablefoot{
            $^{a}$Derived from \textit{Gaia} XP spectra. $^{b}$Derived from our spectra for objects with ($G <$ 14 mag). $^{\rm av}$ The average WR intrinsic color was adopted for calculation as the actual WR subtype is not available (targets not observed in this work or classified as non-WR). \textdagger Candidates with error on \(M_{v}\) greater than 4 mag and not included in Fig.\ref{fig:Abs_Mag_v_all}.
    }

    \vspace{5mm}

{\setlength{\LTcapwidth}{\dimexpr1\textwidth}
\centering

\renewcommand{\arraystretch}{1.4}
\begin{longtable}{cccccccc}
\caption{Absolute M$_{K_s}$ magnitudes of selected WR candidates. The distances from  \citet{2021AJ....161..147B} were used in all the calculations (Table \ref{tab:paral}). See more details in the text.\label{tab:absolute_mags_IR}}\\

\hline\hline
Name & \textit{J} & \textit{H} & $K_s$ & $A_{K_s}^{J}$ & $A_{K_s}^{H}$ & $M_{K_s}$ & Subclass \\
 & {[}mag{]} & {[}mag{]} & {[}mag{]} & {[}mag{]} & {[}mag{]} & {[}mag{]} &  \\
\hline
\endfirsthead
\caption{continued.}\\
\hline\hline
Name & \textit{J} & \textit{H} & $K_s$ & $A_{K_s}^{J}$ & $A_{K_s}^{H}$ & $M_{K_s}$ & Subclass \\
 & {[}mag{]} & {[}mag{]} & {[}mag{]} & {[}mag{]} & {[}mag{]} & {[}mag{]} &  \\
\hline
\endhead
\hline
\endfoot
WR-C-01 & 11.56$\pm$0.03 & 11.06$\pm$0.03 & 10.62$\pm$0.02 & 0.50$\pm$0.07 & 0.66$\pm$0.16 & -3.9$\pm0.4$ & WN4 \\
WR-C-02 & 9.34$\pm$0.02 & 9.03$\pm$0.02 & 8.75$\pm$0.02 & $^{\rm av}$ 0.25$\pm$0.10 &$^{\rm av}$ 0.28$\pm$0.21 & -4.2$\pm0.3$  & Not WR \\
WR-C-03 & 13.44$\pm$0.03 & 13.11$\pm$0.02 & 12.79$\pm$0.03 & $^{\rm av}$ 0.28$\pm$0.11 & 	$^{\rm av}$ 0.34$\pm$0.23 & -1.6$\pm0.7$  & Not WR \\
WR-C-04 & 14.43$\pm$0.03 & 14.11$\pm$0.04 & 13.80$\pm$0.06 & $^{\rm av}$ 0.27$\pm$0.13 & $^{\rm av}$ 0.33$\pm$0.30 & -0.4$\pm0.8$ & .. \\
WR-C-05 & 11.56$\pm$0.03 & 10.94$\pm$0.03 & 10.10$\pm$0.02 & 0.40$\pm$0.06 & 0.36$\pm$0.13 & -5.6$\pm0.7$ & WC5 \\
WR-C-06 & 11.68$\pm$0.02 & 10.33$\pm$0.02 & 8.83$\pm$0.02 & $^{\rm av}$ 1.32$\pm$0.21  &	$^{\rm av}$ 1.99$\pm$0.44 & -6.3$\pm1.1$ & Not WR \\
WR-C-07 & 13.90$\pm$0.04 & 13.39$\pm$0.04 & 13.04$\pm$0.04 & $^{\rm av}$ 0.37$\pm$0.14 & $^{\rm av}$ 0.38$\pm$0.28 & -1.9$\pm0.7$ & .. \\
WR-C-08 & 11.95$\pm$0.02 & 10.71$\pm$0.02 & 9.85$\pm$0.02 & $^{\rm av}$ 0.97$\pm$0.18 & $^{\rm av}$ 1.11$\pm$0.32 & -4.5$\pm1.0$ & .. \\
WR-C-09 & 14.07$\pm$0.04 & 13.69$\pm$0.05 & 13.38$\pm$0.05 & $^{\rm av}$ 0.29$\pm$0.13 & $^{\rm av}$ 0.33$\pm$0.29 & -0.9$\pm0.7$ & .. \\
WR-C-10 & 11.57$\pm$0.05 & 10.74\textdagger & 9.99\textdagger & 0.46$\pm$0.07 & 0.24$\pm$0.03 & -3.9$\pm0.4$ & WC5+abs \\
WR-C-11 & 12.17$\pm$0.03 & 11.57$\pm$0.03 & 11.15$\pm$0.02 & 0.40$\pm$0.06 & 0.38$\pm$0.12 & -4.3$\pm0.5$ & WN5 \\
WR-C-12 & 14.07$\pm$0.02 & 13.54$\pm$0.02 & 13.16$\pm$0.02 & $^{\rm av}$ 0.40$\pm$0.12 & $^{\rm av}$ 0.43$\pm$0.23 & -0.6$\pm0.6$ & .. \\
WR-C-13 & 12.72$\pm$0.06 & 11.67$\pm$0.06 & 10.74$\pm$0.01\textdagger & $^{\rm av}$ 0.74$\pm$0.18 & $^{\rm av}$ 0.73$\pm$0.37 & -3.0$\pm0.8$ & .. \\
WR-C-14 & 10.01$\pm$0.02 & 9.31$\pm$0.03 & 8.76$\pm$0.02 & 0.48$\pm$0.07 & 0.41$\pm$0.12 & -4.3$\pm0.3$ & WC9 \\
WR-C-15 & 12.68$\pm$0.03 & 11.81$\pm$0.03 & 11.25$\pm$0.03 & $^{\rm av}$ 0.65$\pm$0.15 & $^{\rm av}$ 0.68$\pm$0.28 & -3.4$\pm0.8$ & .. \\
WR-C-16 & 12.72$\pm$0.04 & 12.00$\pm$0.04 & 11.29$\pm$0.03 & 0.47$\pm$0.08 & 0.46$\pm$0.17 & -3.2$\pm0.7$ & WC8 \\
WR-C-17 & 15.34$\pm$0.05 & 15.32$\pm$0.09 & 15.16$\pm$0.14 & $^{\rm av}$ 0.05$\pm$0.16 & $^{\rm av}$ 0.12$\pm$0.46 & 0.1$\pm0.9$ & .. \\
WR-C-18 & 9.57$\pm$0.02 & 7.47$\pm$0.02 & 5.85$\pm$0.02 & 1.66$\pm$0.18 & 1.91$\pm$0.31 & -8.8$\pm0.9$ & WC9d \\
WR-C-19 & 12.80$\pm$0.03 & 11.99$\pm$0.03 & 11.18$\pm$0.03 & $^{\rm av}$ 0.57$\pm$0.16 & $^{\rm av}$ 0.56$\pm$0.33 & -4.0$\pm0.8$ & .. \\
WR-C-20 & 10.54$\pm$0.02 & 8.52$\pm$0.06 & 6.94$\pm$0.03 & $^{\rm av}$ 1.51$\pm$0.25 & $^{\rm av}$ 1.65$\pm$0.52 & -3.2$\pm0.6$ & .. \\
WR-C-21 & 11.18$\pm$0.02 & 10.51$\pm$0.02 & 9.69$\pm$0.02 & 0.42$\pm$0.06 & 0.33$\pm$0.10 & -4.2$\pm0.4$ & WC7 \\
WR-C-22 & 11.27$\pm$0.02 & 10.06$\pm$0.02 & 9.13$\pm$0.02 & $^{\rm av}$ 0.81$\pm$0.17 & $^{\rm av}$ 0.73$\pm$0.34 & -5.4$\pm1.1$ & .. \\
WR-C-23 & 10.13$\pm$0.03 & 9.39$\pm$0.03 & 8.91$\pm$0.02 & 0.52$\pm$0.07 & 0.52$\pm$0.14 & -4.6$\pm0.2$ & WN8 \\
WR-C-24 & 10.06$\pm$0.02 & 8.12$\pm$0.02 & 6.48$\pm$0.02 & 1.59$\pm$0.17 & 1.94$\pm$0.32 & -8.3$\pm0.9$ & WC9d \\
WR-C-25 & 12.60$\pm$0.03 & 11.39$\pm$0.02 & 10.33$\pm$0.02 & $^{\rm av}$ 0.88$\pm$0.19 & $^{\rm av}$ 0.92$\pm$0.37 & -4.4$\pm1.3$ & .. \\
WR-C-26 & 9.28$\pm$0.03 & 8.79$\pm$0.03 & 8.39$\pm$0.03 & 0.34$\pm$0.06 & 0.34$\pm$0.13 & -3.8$\pm0.2$ & WN6 \\
WR-C-27 & 12.02$\pm$0.04 & 11.23$\pm$0.04 & 10.66$\pm$0.04 & $^{\rm av}$ 0.62$\pm$0.16 & $^{\rm av}$ 0.69$\pm$0.32 & -3.9$\pm0.8$ & .. \\
WR-C-28 & 8.46$\pm$0.02 & 8.10$\pm$0.02 & 7.84$\pm$0.04 & 0.21$\pm$0.05 & 0.14$\pm$0.10 & -4.9$\pm0.3$ & WN6+abs \\
WR-C-29 & 9.30$\pm$0.03 & 8.91$\pm$0.03 & 8.65$\pm$0.03 & 0.25$\pm$0.05 & 0.22$\pm$0.11 & -3.5$\pm0.2$ & WN7 \\
WR-C-30 & 10.27$\pm$0.02 & 7.93$\pm$0.08 & 6.10$\pm$0.02 & $^{\rm av}$ 1.78$\pm$0.27 & $^{\rm av}$ 2.00$\pm$0.59 & -10.3$\pm1.5$ & .. \\
WR-C-31 & 11.06$\pm$0.03 & 10.08$\pm$0.04 & 9.03$\pm$0.03 & 0.76$\pm$0.10 & 0.95$\pm$0.21 & -5.5$\pm0.6$ & WC8 \\
WR-C-32 & 8.84$\pm$0.03 & 7.23$\pm$0.07 & 5.72$\pm$0.02 & 1.37$\pm$0.16 & 1.77$\pm$0.35 & -8.9$\pm0.8$ & WC9d \\
WR-C-33 & 11.56$\pm$0.02 & 10.24$\pm$0.02 & 9.04$\pm$0.02 & $^{\rm av}$ 0.99$\pm$0.19 & $^{\rm av}$ 1.11$\pm$0.39 & -6.0$\pm1.1$ & .. \\
WR-C-34 & 12.48$\pm$0.03 & 10.80$\pm$0.03 & 9.38$\pm$0.02 & $^{\rm av}$ 1.27$\pm$0.22 & $^{\rm av}$ 1.43$\pm$0.44 & -4.8$\pm1.0$ & .. \\
WR-C-35 & 9.90$\pm$0.03 & 9.02$\pm$0.02 & 8.45$\pm$0.02 & 0.63$\pm$0.08 & 0.65$\pm$0.15 & -5.3$\pm0.4$ & WN8 \\
WR-C-36 & 11.70$\pm$0.05\textdagger & 11.12$\pm$0.05\textdagger & 10.64$\pm$0.03 & 0.52$\pm$0.09 & 0.67$\pm$0.21 & -2.3$\pm0.4$ & WN7h \\
WR-C-37 & 11.40$\pm$0.02 & 10.36$\pm$0.02 & 9.28$\pm$0.02 & 0.90$\pm$0.11 & 1.16$\pm$0.21 & -5.5$\pm0.8$ & WC9 \\
WR-C-38 & 10.36$\pm$0.02 & 9.36$\pm$0.03 & 8.58$\pm$0.02 & 0.74$\pm$0.09 & 0.73$\pm$0.17 & -5.3$\pm0.6$ & WC9 \\
WR-C-39 & 11.60$\pm$0.03 & 10.95$\pm$0.03 & 10.40$\pm$0.03 & 0.51$\pm$0.08 & 0.62$\pm$0.17 & -3.8$\pm0.5$ & WN7-8 \\
WR-C-40 & 9.25$\pm$0.04 & 8.66$\pm$0.04 & 8.26$\pm$0.03 & 0.36$\pm$0.07 & 0.19$\pm$0.11 & -5.2$\pm0.7$ & WC9 \\
WR-C-41 & 12.60$\pm$0.04 & 11.94$\pm$0.03 & 11.44$\pm$0.03 & 0.47$\pm$0.08 & 0.48$\pm$0.15 & -3.6$\pm0.6$ & WN5-6 \\
WR-C-42 & 11.50$\pm$0.03 & 10.78$\pm$0.03 & 10.23$\pm$0.03 & 0.52$\pm$0.08 & 0.56$\pm$0.16 & -4.9$\pm0.6$ & WN6 \\
WR-C-43 & 9.97$\pm$0.03 & 8.11$\pm$0.03 & 6.60$\pm$0.02 & 1.49$\pm$0.17 & 1.76$\pm$0.30 & -7.8$\pm0.8$ & WC9d \\
WR-C-44 & 12.10$\pm$0.03 & 11.41$\pm$0.03 & 10.93$\pm$0.03 & 0.48$\pm$0.08 & 0.45$\pm$0.14 & -3.1$\pm0.5$ & WN5-6 \\
WR-C-45 & 13.29$\pm$0.03 & 12.45$\pm$0.04 & 11.56$\pm$0.03 & $^{\rm av}$ 0.62$\pm$0.16 & $^{\rm av}$ 0.67$\pm$0.36 & -2.2$\pm1.0$ & .. \\
WR-C-46 & 9.98$\pm$0.02 & 9.20$\pm$0.02 & 8.60$\pm$0.03 & 0.46$\pm$0.07 & 0.32$\pm$0.11 & -4.3$\pm0.4$ & WC8 \\
WR-C-47 & 12.53$\pm$0.03 & 11.76$\pm$0.03 & 10.92$\pm$0.03 & 0.47$\pm$0.07 & 0.36$\pm$0.13 & -3.2$\pm0.7$ & WC6 \\
WR-C-48 & 11.49$\pm$0.02\textdagger & 10.76$\pm$0.03\textdagger & 10.06$\pm$0.02\textdagger & 0.39$\pm$0.06 & 0.18$\pm$0.10 & -4.1$\pm0.9$ & WC7 \\
WR-C-49 & 15.64$\pm$0.07 & 15.60$\pm$0.13 & 15.33$\pm$0.17 & $^{\rm av}$ 0.11$\pm$0.19 & $^{\rm av}$ 0.27$\pm$0.58 & - & .. \\
WR-C-50 & 10.66$\pm$0.02 & 10.02$\pm$0.02 & 9.58$\pm$0.02 & 0.43$\pm$0.06 & 0.40$\pm$0.11 & -4.1$\pm0.9$ & WN5-6 \\
WR-C-51 & 13.56$\pm$0.03 & 12.66$\pm$0.04 & 11.99$\pm$0.03 & $^{\rm av}$ 0.71$\pm$0.16 & $^{\rm av}$ 0.83$\pm$0.32 & -3.8$\pm1.0$ & .. \\
WR-C-52 & 13.33$\pm$0.02 & 12.56$\pm$0.03 & 12.01$\pm$0.02 & $^{\rm av}$ 0.59$\pm$0.14 & $^{\rm av}$ 0.66$\pm$0.27 & -2.8$\pm0.7$ & .. \\
WR-C-53 & 14.50$\pm$0.04 & 13.92$\pm$0.05 & 13.39$\pm$0.05 & $^{\rm av}$ 0.49$\pm$0.15 & $^{\rm av}$ 0.63$\pm$0.34 & -1.2$\pm0.9$ & .. \\
WR-C-54 & 8.14$\pm$0.03 & 7.71$\pm$0.02 & 7.46$\pm$0.02 & 0.24$\pm$0.05 & 0.13$\pm$0.07 & -5.4$\pm0.2$ & WN6+abs \\
WR-C-55 & 12.26$\pm$0.02 & 11.57$\pm$0.03 & 10.99$\pm$0.02 & 0.66$\pm$0.08 & 0.85$\pm$0.19 & -4.0$\pm0.6$ & WN3 \\
WR-C-56 & 12.91$\pm$0.02 & 11.84$\pm$0.03 & 11.06$\pm$0.02 & $^{\rm av}$ 0.85$\pm$0.17 & $^{\rm av}$ 0.98$\pm$0.31 & -4.1$\pm1.0$ & .. \\
WR-C-57 & 12.08$\pm$0.03 & 11.16$\pm$0.04 & 10.56$\pm$0.03 & 0.78$\pm$0.10 & 0.89$\pm$0.21 & -4.2$\pm0.6$ & WN/C \\
WR-C-58 & 11.74$\pm$0.02 & 10.95$\pm$0.02 & 10.02$\pm$0.02 & 0.53$\pm$0.07 & 0.49$\pm$0.11 & -4.3$\pm0.4$ & WC5 \\
WR-C-59 & 13.15$\pm$0.04 & 12.72$\pm$0.04 & 12.46$\pm$0.03 & 0.38$\pm$0.07 & 0.41$\pm$0.16 & -2.6$\pm0.5$ & WN3 \\
\end{longtable}}
\tablefoot{
            $^{\rm av}$ The average WC or WN intrinsic colors were adopted for calculation (based on \textit{Gaia} ELS subtype) as the actual WR subtype is not available (targets not observed in this work or classified as non-WR). \textdagger The quality flag indicates possible data issues with the particular 2MASS measurements.
    }
\end{appendix}

\end{document}